\documentclass{cernyrep}

\usepackage{graphicx}
\usepackage{dcolumn}
\usepackage{bm}
\usepackage[bookmarks, colorlinks=true, linktoc=page, pdftex, linkcolor=black, citecolor=black, urlcolor=blue]{hyperref}
\usepackage[mathlines]{lineno}
\usepackage{upgreek}
\usepackage{booktabs}
\usepackage{amsmath}
\usepackage{textcomp}
\usepackage{url}
\usepackage{fancyhdr}
\usepackage[colorlinks=true, linkcolor=black, citecolor=black, urlcolor=blue]{hyperref}

\fancyhfoffset{4 mm}
\fancypagestyle{ARTTITLE}{%
\fancyhf{} 
\lhead{\small{Proceedings of the CERN--Accelerator--School course: \it{Introduction to Accelerator Physics}}}
\lfoot{Available online at \url{https://cas.web.cern.ch/previous-schools}}
\rfoot{\thepage\hspace*{3mm}}

}

\begin{document}
\title{Synchrotron light sources and X-ray free-electron-lasers}
\author{Eduard Prat}
\institute{Paul Scherrer Institute, Villigen, Switzerland}

\begin{abstract}
Synchrotron light sources and X-ray free-electron laser (FEL) facilities are unique tools providing extremely brilliant X-rays that allow the observation of matter with atomic spatial resolution.  
On the one hand, synchrotron light sources consist of electron circular accelerators and produce synchrotron radiation in bending magnets and undulators. 
On the other hand, X-ray FEL facilities are based on electron linear accelerators and generate more coherent and shorter pulses suitable for time-resolved experiments. 
In this contribution we will qualitatively describe synchrotron and X-ray FEL facilities. We will start explaining some fundamental concepts related to synchrotron and FEL radiation. We will then describe the two kinds of machines, including the history and current facilities, the typical layout, and some basic concepts about the electron beam dynamics and properties. 

\end{abstract}


\maketitle

\section{From synchrotron radiation to free-electron lasers}
In this chapter we will qualitatively describe some basic concepts about synchrotron radiation emitted in bending or dipole magnets, undulator radiation, and radiation generated under the free-electron laser (FEL) process. 
We recommend the following references for further and deeper reading: the books on synchrotron radiation and FELs by Kim, Huang and Lindberg \cite{Kim17} and by Willmott~\cite{Wil19}, the chapter on synchrotron radiation of Wiedemann's book~\cite{Wie15}, the article on synchrotron and undulator radiation by Kim~\cite{Kim89}, the book focused on undulators by Clarke~\cite{Cla04}, and the review article on FELs by Pellegrini, Marinelli and Reiche~\cite{Pel16}.

\subsection{Synchrotron radiation}
This section has been inspired by the chapter about synchrotron radiation of the book of Wille~\cite{Wil01}. 
Accelerated charged particles emit electromagnetic radiation following Maxwell equations. 
In the case of radially accelerated charges, the associated radiation is called \emph{synchrotron radiation}.
This phenomenon occurs in bending magnets and was first observed in synchrotron facilities, where the beam energy and magnet dipole strengths are ramped up \emph{synchronously} -- hence, the name synchrotron radiation.

\subsubsection{Radiation power of a charged particle}

The total power radiated by a non-relativistic particle as it accelerates can be calculated using the Larmor formula~\cite{Lar97}:
\begin{equation}
P = \left(\frac{e}{mc^2}\right)^2\frac{c}{6\pi\varepsilon_0}\left(\frac{d\vec{p}}{dt}\right)^2 = \frac{e^2a^2}{6\pi\varepsilon_0c^3}, \label{eq:larmor}  
\end{equation}
where $e$ is the charge of the particle, $m$ is the particle's rest mass, $c$ is the speed of light, $\varepsilon_0$ is the vacuum permittivity, $\vec{p}$ is the momentum of the particle, $t$ is time, and $a$ is the particle's acceleration.

We will now generalize the above formula for relativistic particles. For that we apply an invariant formulation by exchanging the momentum $\vec{p}$ with the four-momentum $P_\mu$. 
The four-momentum is expressed as $P_\mu = (E_e/c,\vec{p})$, where $E_e$ is the energy of the particle. 
In Eq.~\ref{eq:larmor} we replace $\left(\frac{d\vec{p}}{dt}\right)^2$ with:
\begin{equation}
\left(\frac{d P_\mu}{d\tau}\right)^2 = \left(\frac{d\vec{p}}{d\tau}\right)^2 - \frac{1}{c^2}\left(\frac{d E_e}{d\tau}\right)^2, 
\end{equation} 
with $\tau=\frac{1}{\gamma}t$ being the time in the moving system, where $\gamma$ is the Lorentz factor ($\gamma=E_e/(mc^2)$).
Note that in the above equation the energy term has a negative sign, which arises from calculating the invariant in special relativity.  

By doing that we obtain the radiation power of a charged particle $P$: 
\begin{equation}
P = \left(\frac{e}{mc^2}\right)^2\frac{c}{6\pi\varepsilon_0}\left[\left(\frac{d\vec{p}}{d\tau}\right)^2 -  \frac{1}{c^2}\left(\frac{dE_e}{d\tau}\right)^2\right].  
\label{eq:larmor_tot} 
\end{equation}
We will distinguish two cases, depending on whether the acceleration is linear (i.e. parallel to the momentum of the particle) or circular (i.e. perpendicular to the particle's momentum).

\subsubsubsection{Linear acceleration}
In the linear acceleration case the acceleration $d\vec{p}/dt$ is parallel to the momentum of the particle $\vec{p}$ and the energy of the particle is changed. Considering that $E_e^2 = (mc^2)^2 + p^2c^2$, we can write: 
\begin{equation}
E_e\frac{dE_e}{d\tau} = c^2p\frac{dp}{d\tau}.
\end{equation}
From the above expression we can derive the energy variation as
\begin{equation}
\frac{dE_e}{d\tau} = v\frac{dp}{d\tau},
\end{equation}
where we have used $p=mv$, with $v$ being the velocity of the particle. 
We can replace the previous expression in Eq.~\ref{eq:larmor_tot}. After some algebra and substituting $\tau$ to $t$ again, we find that:
\begin{equation}
P = \left(\frac{e}{mc^2}\right)^2\frac{c}{6\pi\varepsilon_0}\left(\frac{dp}{dt}\right)^2 = \left(\frac{e}{mc^2}\right)^2\frac{c\beta^2}{6\pi\varepsilon_0}\left(\frac{dE_e}{dz}\right)^2, 
\end{equation}
where  $\beta=v/c$ is the relative velocity of the particle with respect to the speed of light and $z$ is the direction of acceleration. 
Taking a numerical example for realistic acceleration parameters, for a gradient $dE_e/dz$ of 15 MeV/m, the total radiated power for an electron would be $4\times10^{-17}$~W. 
From this example we can conclude that the radiation power associated to linear acceleration is negligible. 

\subsubsubsection{Circular acceleration}
In this case, the acceleration is perpendicular to the particle's momentum and the energy is constant, i.e. $dE_e/d\tau=0$. The acceleration can be expressed as 
\begin{equation}
\frac{dp}{dt} = \frac{m\gamma v^2}{R} = \frac{\beta^2E_e}{R},
\end{equation}
where $R$ is the radius. 
We can derive the equation for the radiated power from Eq.~\ref{eq:larmor_tot} considering that $\gamma=E_e/mc^2$ and after some algebraic operations:
\begin{equation}
P = \frac{e^2c}{6\pi\varepsilon_0}\left(\frac{E_e}{mc^2}\right)^4\frac{\beta^4}{R^2}.
\label{eq_Psync}
\end{equation}
We observe that the power of a particle is proportional to the fourth power of the energy and inversely proportional to the fourth power of the mass: $P\propto (E_e/m)^4$. 
As we will see in the following, the power associated to circular acceleration, called synchrotron radiation, is not negligible and in fact can be used as a powerful tool to investigate matter. 

\subsubsection{Energy loss per turn due to synchrotron radiation in a circular accelerator}
The energy loss per turn in a circular accelerator $U_0$ can be calculated multiplying the power (Eq.~\ref{eq_Psync}) by the required time for a particle to make a turn: 
\begin{equation}
U_0 = Pt_\text{turn} = P\frac{2\pi R}{v}.
\end{equation}
Assuming relativistic particles ($\beta \approx 1$), the above expression can be expressed in practical units as follows: 
\begin{equation}
U_0(\text{keV}) = 10^{33}\frac{e}{3\varepsilon_0}\left(\frac{e}{mc^2}\right)^4\frac{[E_e(\text{GeV})]^4}{R(\text{m})}.
\end{equation}
The first term of the above equation $10^{33}\frac{e}{3\varepsilon_0}\left(\frac{e}{mc^2}\right)^4$ is approximately 88.5 for electrons and $7.8\times10^{-12}$ for protons. In other words, due to its lower mass, an electron loses about $1\times10^{13}$ more synchrotron radiation power than a proton. 
This fact explains why synchrotron light sources utilize electron and not proton beams.

We can calculate the energy loss per turn and required power for some real accelerators: 
\begin{itemize}
\item Swiss Light Source (SLS) in Villigen (Switzerland). With an electron energy of 2.4~GeV and a bending radius of 5.7~m, the energy loss per turn is $U_0=515$~keV. For an electron beam current $I$ of 400~mA, the power loss per turn due to synchrotron radiation is $U_0 I=206$~kW. This is not negligible and has to be resupplied by the RF system to keep the same electron beam energy. 
\item LEP-II in Geneva (Switzerland). The electron beam energy was 100~GeV, with a bending radius of 3026~m and a current of 6~mA. This corresponds to an energy loss per turn of 2.9~GeV and a power of 17.5~MW. 
\item LHC in Geneva (Switzerland). In this case we have protons of 7~TeV, a radius of 2804~m, and a current of 580~mA, corresponding to an energy loss per turn of 6.7~keV and a power of 3.9~kW. 
\end{itemize}

From these examples we understand that electron circular accelerators are limited by the RF power that needs to be supplied to compensate the energy loss due to the emission of synchrotron radiation. 

\subsubsection{Properties of synchrotron radiation}

We start by introducing the Planck-Einstein relation that links the photon energy $E$, the photon frequency $\nu$, and the photon wavelength $\lambda$:
\begin{equation}
E = h\nu = \frac{hc}{\lambda},
\end{equation}
where $h$ is the Planck constant ($h=6.626\times10^{-34}$~m$^2$kg/s).

\begin{figure}[ht]
\begin{center}
\includegraphics[width=0.9\linewidth]{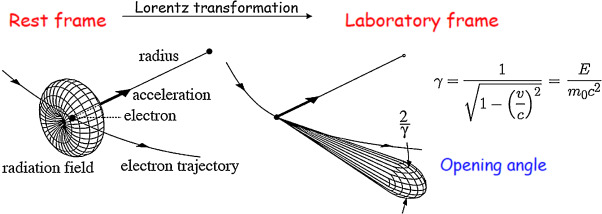}
\caption{Emission pattern of the radiation of a relativistic charged particle within its own frame of motion (left) and as observed in the laboratory frame (right). Image taken from~\cite{Ebe15}. }
\label{fig:col}
\end{center}
\end{figure}

\subsubsubsection{Angular distribution of emission}
A charged particle that is centripetally accelerated will emit radiation mostly in the directions perpendicular to acceleration.
The radiation emission pattern follows a characteristic donut shape in the rest frame, as shown on the left side of Fig.~\ref{fig:col}. 

Due to the Lorentz transformation from the rest frame to the laboratory system, the donut shape emission is heavily distorted towards forward emission with an opening angle of $1/\gamma$. 
For an estimate we consider one photon emitted perpendicular to the direction of acceleration and motion. Considering that the particle is accelerated in $x$ and moves in $z$, in the rest system this photon has a momentum $p_{y'} = E'/c$. 
Now we do a Lorentz transformation to the laboratory system to obtain: $p_y = p_{y'}$, $p_z = \gamma p_{y'}$.  
The emission angle in the laboratory frame will be $p_y/p_z = 1/\gamma$ (see right side of Fig.~\ref{fig:col}). 
Thus, for relativistic particles the radiation will be strongly collimated. 
As an example, for the European Synchrotron Radiation Facility (ESRF) in Grenoble (France), which has an electron energy of 6~GeV, the emission angle is 85~\textmu rad, equivalent to a beam spot of 1~cm at a location 60~m from the source point.

\begin{figure}[ht]
\begin{center}
\includegraphics[width=0.5\linewidth]{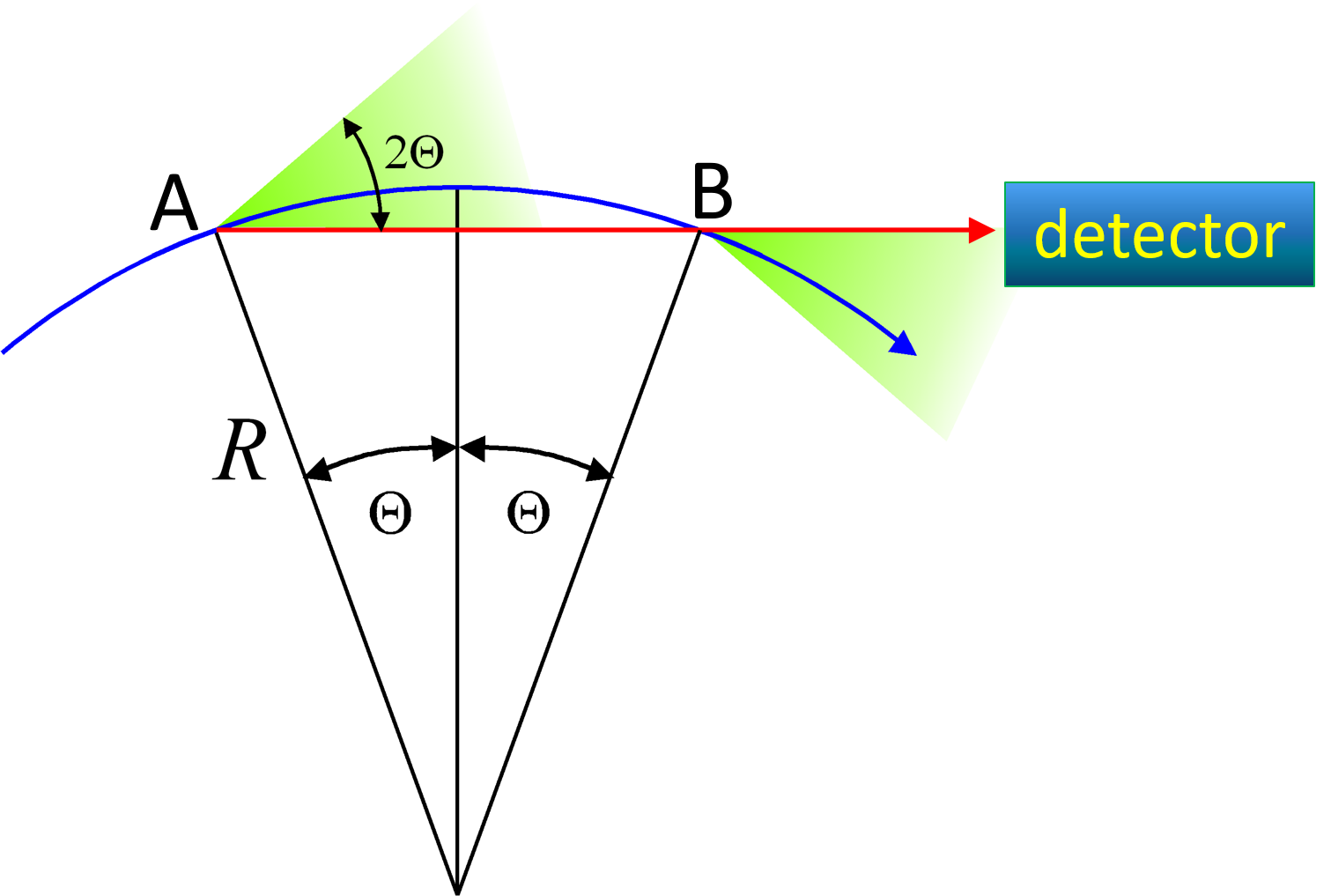}
\caption{Sketch to show how to calculate the time that a sample will be illuminated in a detector. Image courtesy: Andreas Streun.}
\label{fig:time}
\end{center}
\end{figure}

\subsubsubsection{Time structure}
From the opening angle seen in the previous subsection we can estimate how long an experimental sample will be illuminated in a detector station. To explain this we will use the sketch in Fig.~\ref{fig:time}. 
The time difference between the first photon (emitted at point $A$ in the figure) and the last photon (emitted at point $B$) corresponds to the time delay between the electron and the photons:
\begin{equation}
\Delta t = \frac{2R\theta}{c\beta}  - \frac{2R\sin{\theta}}{c}.
\end{equation} 
Considering that the opening angle is $\theta=1/\gamma$, using $\sin{\theta} \approx \theta - \theta^3/6$ (approximation valid for small angles), and considering that $\frac{1}{\beta} = \frac{1}{\sqrt{1-1/\gamma^2}} \approx 1+\frac{1}{2\gamma^2}$ (approximation valid for large $\gamma$), we can get the pulse duration as
\begin{equation}
\Delta t \approx \frac{4R}{3c\gamma^3}. 
\end{equation}

The typical frequency associated to this time interval is $\nu_\text{typ} = \frac{1}{\Delta t}$. The typical photon energy is: 

\begin{equation}
E_\text{typ} = h\nu_\text{typ}  \approx \frac{3ch\gamma^3}{4R}.
\end{equation} 
For the example at the ESRF, the electron beam energy is 6~GeV and the bending radius is 23~m, corresponding to $\Delta t=6.3\times10^{-20}$~s and $E_{text{typ}}=65$~keV.

\begin{figure}[ht]
\begin{center}
\includegraphics[width=0.6\linewidth]{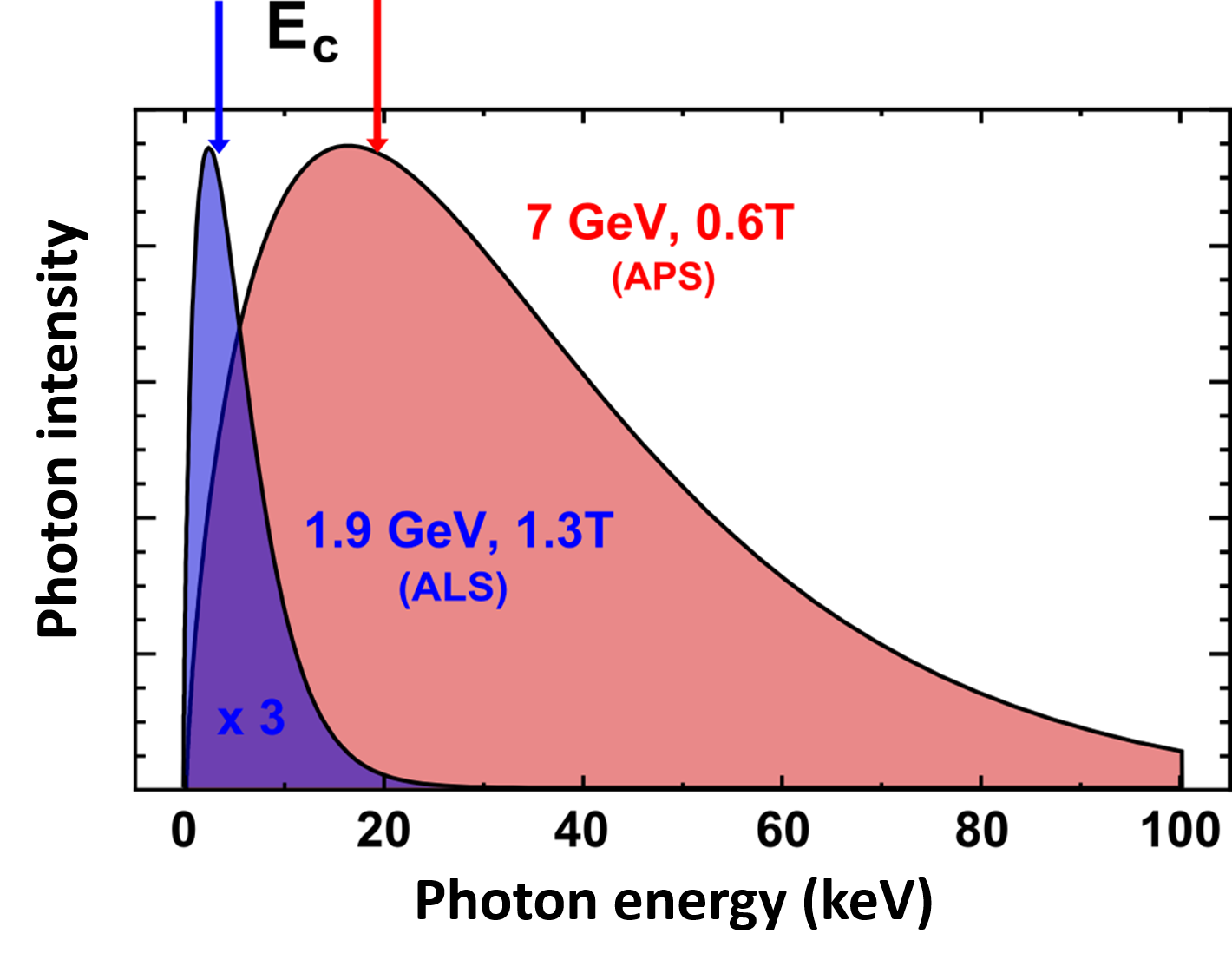}
\caption{Radiation spectrum and critical energy for two examples of real accelerators in the USA: the Advanced Photon Source (APS), with an electron energy of 7~GeV and a magnetic field of 0.6~T, and the Advanced Light Source (ALS), with an energy of 1.9~GeV and a field of 1.3~T. Original source: \cite{PSDESY}. }
\label{fig:spectrum}
\end{center}
\end{figure}

\subsubsubsection{Radiation spectrum and critical energy}
The exact radiation spectrum of synchrotron radiation follows a modified Bessel function first derived by Schwinger~\cite{Sch49}.
The critical photon energy $E_c$ divides the spectrum into two halves of equal integrated power. It is related to the typical energy by
\begin{equation}
E_c = \frac{E_\text{typ}}{\pi} \approx \frac{3c\hbar\gamma^3}{2R},
\label{eq:Ec}
\end{equation}
where $\hbar=h/(2\pi)$ is the reduced Planck constant. 
A practical formula is
\begin{equation}
E_c\text{(keV)} \approx 0.665B\text{(T)}E^2\text{(GeV$^2$)}.
\label{eq:Ec2}
\end{equation}
For the ESRF example, $E_e=6$~GeV and $R=23$~m correspond to a magnetic field $B$ of 0.87~T, for which we obtain a critical energy of 21~keV. 
Figure~\ref{fig:spectrum} shows the typical radiation spectrum and the critical energy for two additional examples.

\subsection{Undulator radiation}

Up to now we have seen the synchrotron radiation emitted in a bending or dipole magnet. 
In this section we will consider the radiation produced in \emph{undulators}, devices which consist of a periodic structure of dipole magnets with alternating polarity. 
The periodicity of the undulator is defined by the number of periods or bending magnets $N$ and by the undulator period $\lambda_{u}$, the latter with typical values of a few centimeters. 
The radiation emitted in undulators has higher power and better quality than the radiation emitted in an individual bending magnet. 
One of the main advantages of undulators is that the deflection alternates so that the global electron trajectory is straight (in contrast to the curved trajectory in bending magnets), allowing an increase of the radiation flux at the location of the experimental station scaling with the length of the undulator.

\begin{figure}[ht]
\begin{center}
\includegraphics[width=0.5\linewidth]{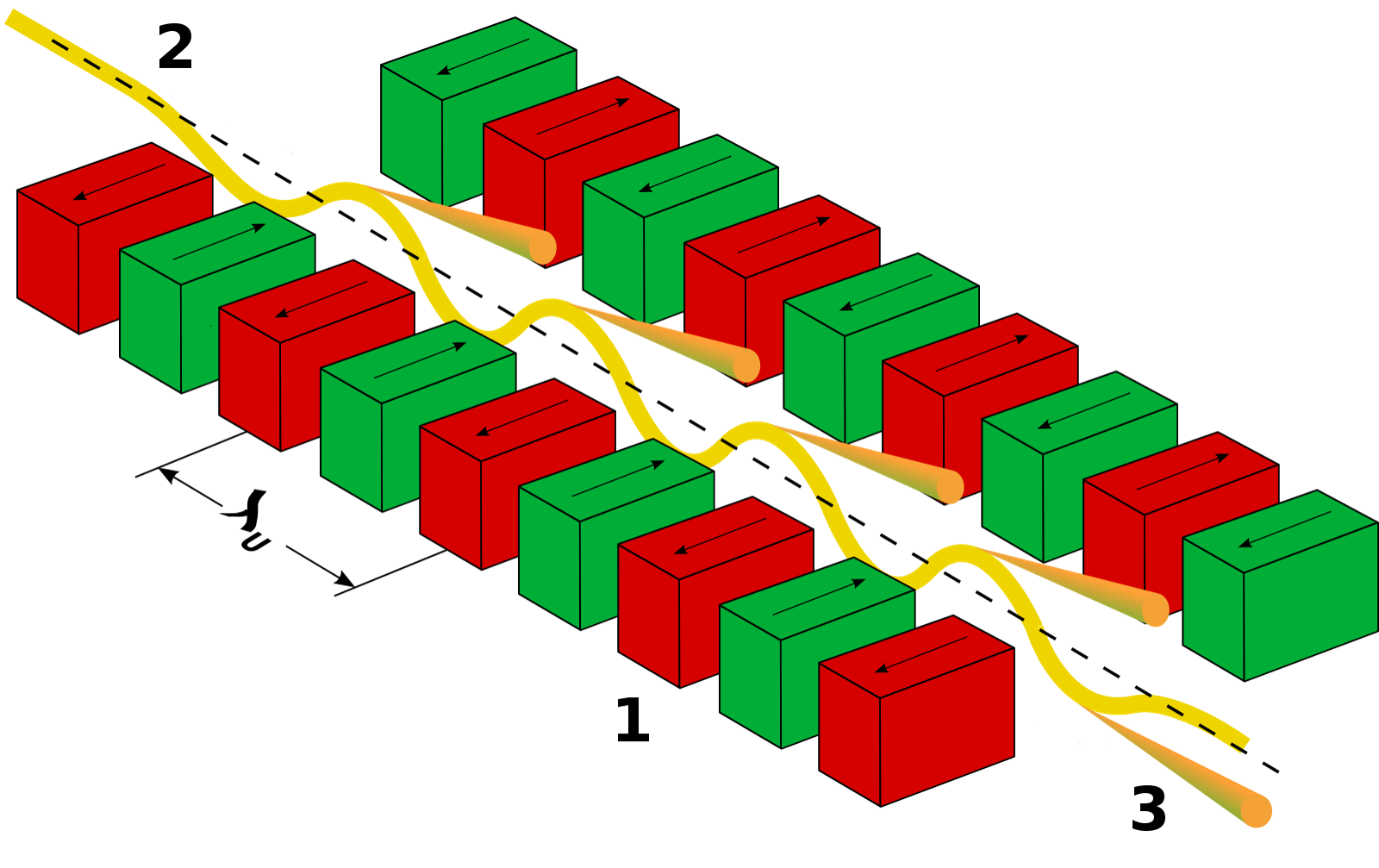}
\caption{Sketch of an undulator and its working principle. 1: magnets with alternating polarity, 2: the yellow line indicates the electron beam trajectory entering from the upper left; 3: the orange cones show the synchrotron radiation produced in the forward direction. Image from Bastian Holst~\cite{und_imag}.}
\label{fig:undulator}
\end{center}
\end{figure}

There are mainly two types of undulators depending on whether the magnetic field is in one or two directions: planar, with magnetic field components in one transverse direction resulting in electron trajectories also restricted to one plane; and helical, with magnetic field components in two directions resulting in a helical trajectory of the electrons. 
Here we will limit ourselves to planar undulators. 
Figure~\ref{fig:undulator} shows a sketch of an undulator and its working principle.

The on-axis magnetic field of a standard planar undulator is non-zero only in the vertical (horizontal) direction, assuming the motion of the electron beam is in the horizontal (vertical) plane, and it can be expressed as
\begin{equation}
B_y=B_0\cos{k_u z},
\label{eq:undB}
\end{equation}
where $B_0$ is the peak magnetic field, $k_u=2\pi/\lambda_u$ is the wave number of the undulator, and $z$ is the longitudinal coordinate along the undulator. 
The undulator field is normally described by the undulator parameter $K$:
\begin{equation}
K = \frac{eB_0}{mck_u}.
\end{equation}
$K$ is dimensionless and has typical values around 1. 
In practical units it can be expressed as $K \approx 0.93 B_0(\text{T})\lambda_u(\text{cm})$.

\subsubsection{Motion in the undulator}
The motion of an electron traveling through an undulator is defined by the Lorentz force: $\vec{F} = e(\vec{v}\times\vec{B})$. 
Here, $\vec{v}$ is the velocity of the electron, which is dominant in the longitudinal direction $z$ and can be approximated as $v_z=\beta_z c$. Assuming that the magnetic field $\vec{B}$ is dominant in the vertical plane $y$, the force will be dominant in the horizontal direction $x$ (see Eq.~\ref{eq:undB}): 
\begin{equation}
F_x = e(-v_z B_y) \approx -ec\beta_z  B_0\cos{k_u z}.
\end{equation}
We know that the force is the derivative of the momentum over time, so we can write, considering that $\gamma^2=1/(1-\beta^2)$ is constant (since the energy is conserved during motion under the influence of a magnetic field) that
\begin{equation}
F_x = \frac{dp_x}{dt} = \gamma mc \frac{d\beta_x}{dt}.
\end{equation}

From the two previous equations we can derive, after some algebra, the transverse and longitudinal motion of the electrons in the undulator:
\begin{gather}
\beta_x = - \frac{K}{\gamma}\sin{k_u z}, \\
 \beta_z = 1 - \frac{1+K^2/2}{2\gamma^2} + \frac{K^2}{4\gamma^2}\cos(2k_u z) = \langle \beta_z \rangle + \frac{K^2}{4\gamma^2}\cos(2k_u z), 
\label{eq:bz}
\end{gather}
where $\langle \beta_z \rangle$ is the average longitudinal velocity over one undulator period relative to the speed of light:
We observe that the longitudinal wiggle has twice the period than the transverse one.
In the rest frame of the undulator, this results in an electron trajectory with a characteristic 8 shape.

\subsubsection{Resonance condition}
\begin{figure}[ht]
\begin{center}
\includegraphics[width=0.6\linewidth]{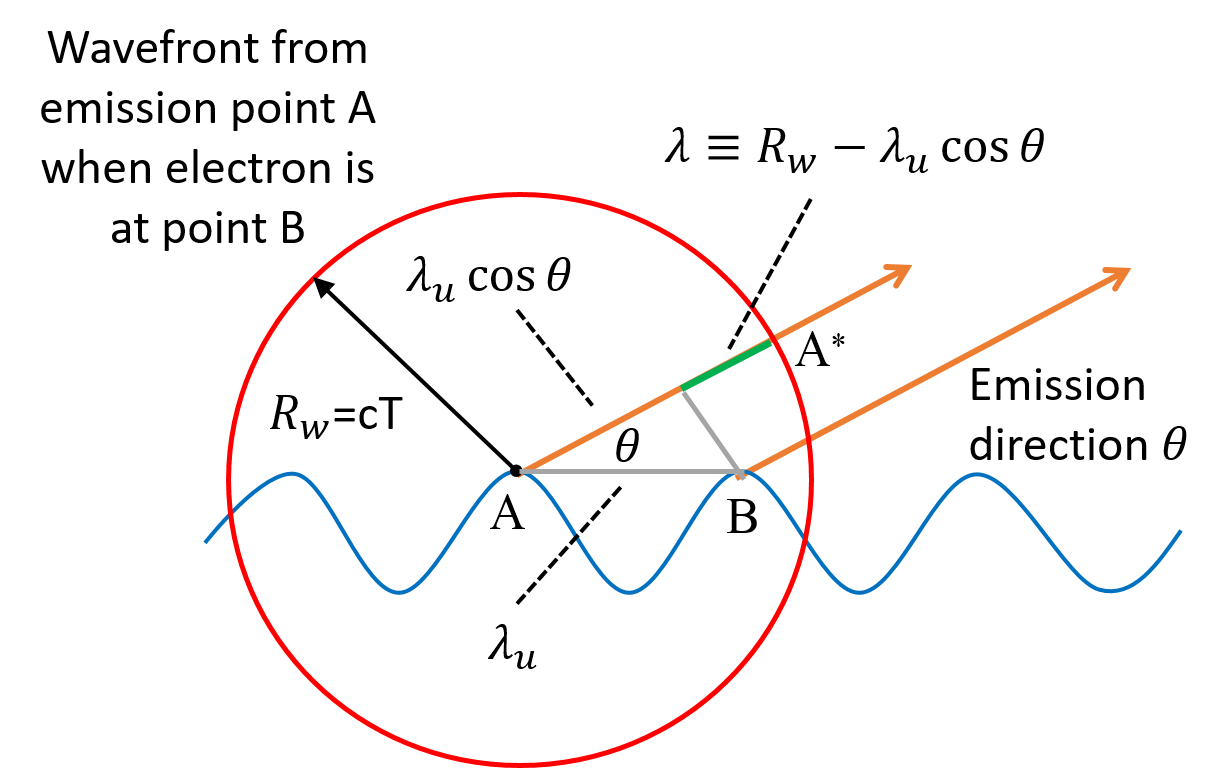}
\caption{Sketch to show the condition for constructive interference between photons emitted at different locations in the undulator. The blue line shows the trajectory of the electron within the undulator. At point $A$ the electron emits radiation in all directions. 
When the electron has moved by one undulator period and it is at point $B$, the wavefront emitted from $A$ has a radius of $R_w$. This wavefront is indicated by the red circle, with the point $A^*$ indicating the emission in the direction $\theta$. 
At that moment the electron emits an identical wavefront from point $B$. 
After certain time (i.e., in the far field), the distance between the two wavefronts in the direction $\theta$ settles to a constant value indicated by the length of the green line in the sketch. 
For a constructive interference between both wavefronts this distance has to be exactly the radiation wavelength $\lambda$ (or an integer multiple $n$ of it); i.e.,  $n\lambda=R_w -\lambda_u\cos{\theta}$. Image courtesy of Sven Reiche.}
\label{fig:resonance}
\end{center}
\end{figure}

Figure~\ref{fig:resonance} shows a sketch useful to understand how to have constructive interference  between the radiation emitted by the same electron at different locations in the undulator. 
To achieve this constructive interference, the electron must slip back by exactly one radiation wavelength $\lambda$ (or an integer multiple of it) over one undulator period. 
In order to fulfill this condition, under an arbitrary emission direction $\theta$, the radiation wavelength (or an integer multiple $n$ of it) has to be exactly $R_w -\lambda_u\cos{\theta}$, where $R_w$ is the radius of the wavefront emitted by the electron one undulator period before. 

The value of the radius $R_w$ can be obtained as $R_w=cT$, where $T$ is the time needed for the electron to move by one undulator period; i.e. 
\begin{equation}
T = \frac{\lambda_u}{\langle \beta_z \rangle c}.
\end{equation}
From this and the expression of $\langle \beta_z \rangle$ in Eq.~\ref{eq:bz}, we have: 
\begin{equation}
R_w=cT = c \frac{\lambda_u}{\langle \beta_z \rangle c} = \lambda_u \frac{1}{1-\frac{1+K^2/2}{2\gamma^2}} \approx \lambda_u\left(1+\frac{1+K^2/2}{2\gamma^2}\right).
\end{equation}
Moreover, for a small emission angle $\theta$, we can write $\lambda_u\cos{\theta}\approx\lambda_u(1-\theta^2/2)$. 
If we put this together with $n\lambda = R_w -\lambda_u\cos{\theta}$ we find the following condition: 

\begin{equation}
\lambda = \frac{\lambda_u}{2n\gamma^2}\left(1+\frac{K^2}{2} + \theta^2\gamma^2\right). 
\label{eq:lam0}
\end{equation}
In the forward direction ($\theta=0$) we obtain the so-called resonance condition: 
\begin{equation}
\lambda_R = \frac{\lambda_u}{2n\gamma^2}\left(1+\frac{K^2}{2}\right). 
\label{eq:lam}
\end{equation}
We observe that the wavelength increases with $\theta^2$, i.e., it  gets longer as we move away from the axis. The resonance condition shows that the radiation wavelengths become shorter than the undulator period by a factor proportional to $1/\gamma^2$. 
We can also observe that the radiation wavelength increases with the magnetic field, parameterized by $K$. 
Furthermore we note that the resonant photon wavelength or energy is different from the critical photon energy (see Eq.~\ref{eq:Ec}) and that the critical photon energy defines how high the harmonic content can be in the undulator radiation spectrum.

An important implication of the resonance condition is that the undulator radiation wavelength can be tuned: it can be controlled by either changing the electron beam energy ($E_e\propto\gamma$) or by varying the undulator parameters (the period $\lambda_u$ or the magnetic field $K$). 
As an example for the SwissFEL case and for the fundamental radiation ($n=1$), an undulator period of 15~mm, a $K$-value of 1.2 and an energy of 5.8~GeV would give a radiation wavelength of 0.1~nm. 
We note that the wavelength of the synchrotron radiation emitted in bending magnets can also be tuned with the electron beam energy (as evidenced in Eq.~\ref{eq:Ec} and Fig. \ref{fig:spectrum}).

\subsubsection{Properties of undulators and wigglers}

Historically, periodic structures of dipoles with alternating polarity were called either undulators or wigglers, depending on their
field strength: devices resulting in large deflections of the electrons ($K>1$) were called `wigglers,' whereas the term `undulators' was
applied to structures giving rise to only small trajectory angles ($K \leq 1$).
The distinction made sense back then, since the consistency with which the dipoles could be manufactured was still
limited at the time: The relatively poor achievable field quality had only a limited effect at small fields ($K \leq 1$) whereas it mattered
a lot at larger field strengths ($K>1$).
Nowadays modern fabrication techniques provide sufficient field quality even for large $K$ values, and the distinction no longer makes sense from a physics point of view. The two types of devices, undulators and wigglers, are collectively called \text{insertion devices}.

The harmonic content of the undulator radiation depends on the field strength $K$: we see from Eq.~\ref{eq:Ec2} and Fig.~\ref{fig:spectrum}
how the achievable photon energies increase with the available magnetic field.  
For $K$ values on the order of one the spectral output of the radiation is confined to a few harmonic lines, with peaks being produced at the harmonics of the resonant frequency. 
With growing $K$ values the harmonic content increases. The effective cutoff for the harmonic generation is given by the critical energy of the individual dipoles of the insertion device (Eq.~\ref{eq:Ec}). 

Assuming an undulator with $N$ periods, an electron will create a radiation pulse with length equal to $N$ times the fundamental radiation wavelength (because of the slippage). Due to the longer pulse, the spectral width will decrease proportionally to the number of periods. The relative bandwidth in an undulator of $N$ periods can be approximated as
\begin{equation}
\frac{\Delta \lambda}{\lambda} \approx \frac{1}{Nn} .
\label{eq:delam}
\end{equation}
For example, an undulator with 100 periods will produce radiation with a relative bandwidth of about 1\% at the fundamental wavelength.

In practice, the bandwidth is limited by the intrinsic broadening due to the field quality. Based on Eq.~\ref{eq:lam}, the relative bandwidth associated to a variation in undulator field $\Delta K$ is
\begin{equation}
\frac{\Delta \lambda}{\lambda}\approx \frac{2\Delta K K}{2+K^2}.
\label{eq222}
\end{equation}
For large $K$ values this reduces to $\Delta \lambda/\lambda\approx2(\Delta K/K)$, whereas for small $K$ values the approximation
$\Delta \lambda/\lambda\approx K^2(\Delta K/K)$ holds. 
From these approximations we can understand the historical distinction between undulators ($K<1$) and wigglers ($K>1$), since the former could provide smaller bandwidth broadening with the same field error $\Delta K/K$.

Because of the bandwidth reduction (see Eq.~\ref{eq:delam}) and the increase in radiated power, both proportional to the number of bending magnets $N$, the spectral power in an undulator will increase as $N^2$. This is true as long as the bandwidth is not limited by field errors (see Eq.~\ref{eq222}). In this case, the spectral power will only increase proportionally to $N$. Again, this was historically one of the main reasons to distinguish between wigglers ($K>1$, spectral power proportional to $N$) and undulators ($K \leq 1$, spectral power proportional to $N^2$). 

The above discussion assumed a fixed emission angle $\theta$, which would be equivalent to observing the radiation after passing a small pinhole. If we consider all allowed emission angles there will be an effective spectral broadening according to Eq.~\ref{eq:lam0}. This can give rise to overlaps in the harmonic content such that the spectrum appears continuous. 
A measure of the acceptable opening angle of the emission to limit this spectral broadening is given by the so-called coherence angle $\theta_c$. It is found by equating the spectral width due to slippage and field quality (Eqs.~\ref{eq:delam} and \ref{eq222}) to the spectral broadening due to off-axis emission given by Eq.~\ref{eq:lam0}). In the case of negligible impact due to field quality we find (with $\frac{\Delta \lambda}{\lambda}= \frac{\lambda(\theta_c)-\lambda(0)}{\lambda(0)}$): 
\begin{equation} 
\theta_c=\frac{\sqrt{\frac{\Delta \lambda}{\lambda} (1+\frac{K^2}{2})}}{\gamma} \approx\frac{\sqrt{\frac{1}{Nn}(1+\frac{K^2}{2})}}{\gamma}.
\label{eq:thetac}
\end{equation}
The coherence angle should not be confused with the general opening angle, which includes all possible frequencies the undulator can produce (as in a bending magnet). 

Two important special cases can be derived from Eq.~\ref{eq:thetac}: for 
a small $K$ value the coherence angle becomes
\begin{equation}
\theta_c = \frac{1}{\sqrt{Nn}\gamma},
\end{equation} 
or, for the fundamental radiation ($n=1$), simply 
\begin{equation}
\theta_c = \frac{1}{\sqrt{N}\gamma},
\label{eq333}
\end{equation} 
which means that in an undulator with small $K$, the emission angle for a given frequency is reduced by a factor of $\sqrt{N}$ with respect to the opening angle of bending-magnet radiation. 
The other special case concerns wigglers with both large $K$ and large variation in $K$ (therefore $\Delta\lambda/\lambda \approx 1$), such that, for the fundamental radiation, 
\begin{equation}
\theta_c = \frac{K}{\gamma}.
\label{eq444}
\end{equation}
For this particular case, the coherence angle coincides with the general opening angle. This means that, unlike in the undulator case and similar to a dipole magnet, all frequencies are produced at all emission angles. 

\begin{figure}[ht]
\begin{center}
\includegraphics[width=1\linewidth]{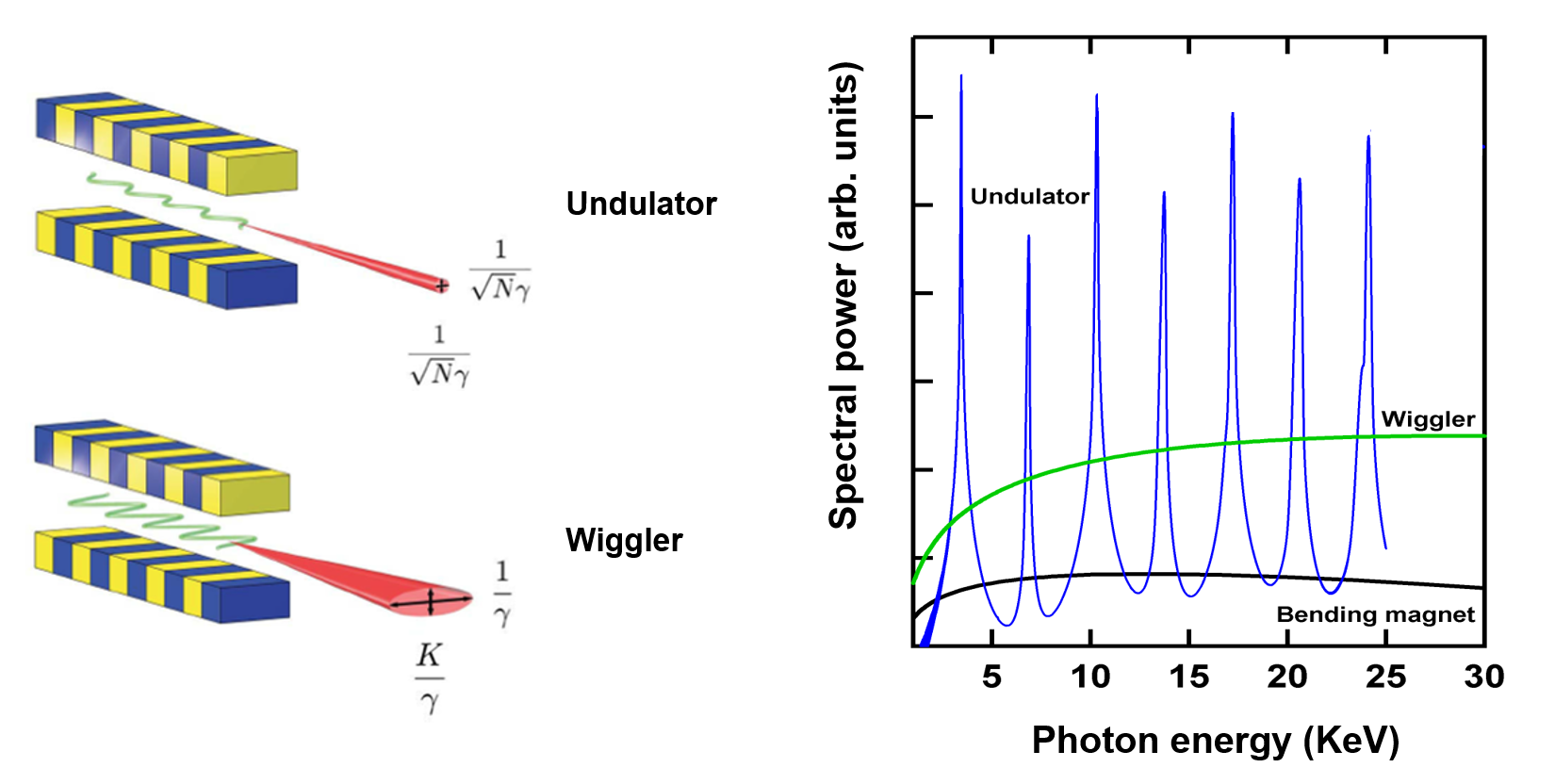}
\caption{Comparison between the radiation properties of wigglers and undulators. Left: cartoon showing the radiation cones defined by the coherence angles (original source: \cite{Nas_lecture}). Right: spectral power as a function of the photon energy (original source: \cite{PSDESY}). See text for more details. }
\label{fig:und_wig}
\end{center}
\end{figure}

Figure~\ref{fig:und_wig} illustrates the differences in emission cones and spectral properties between undulators and wigglers. The emission cones are defined by the coherence angles. For that, we assume that the undulator has a small $K$ value (Eq.~\ref{eq333}) and that the wiggler has a large $K$ with substantial variation $\Delta K$ (Eq.~\ref{eq444}). 
The smooth curve illustrating the spectral properties of the wiggler radiation is the result of
overlapping harmonics, including broadenings associated both with field quality (relevant for higher harmonics)
as well as emission angles $\theta$ larger than the coherence angle $\theta_c$ (relevant for lower harmonics). We include, for the sake of comparison, the spectrum of a bending magnet.

\subsection{Free-electron lasers}
Having described the synchrotron radiation produced in bending magnets and in undulators, we will now explain how an electron beam traveling through an undulator beamline generates radiation following the free-electron laser (FEL) process. 
In an FEL, there is not only a coherent contribution of the radiation produced in each undulator period, but also a coherent addition of the radiation produced by each electron. 
In this case, the coherence is related to the formation of particle distributions with dimensions smaller than the resonant wavelength (called microbunching). 
Because of this, the radiation power is proportional to the square number of electrons within the cooperation length of the FEL process. 
This implies that FEL radiation power is orders of magnitude higher than the power of standard undulator radiation.

\subsubsection{Energy modulation, microbunching, and enhanced radiation emission}
The FEL is a circular runaway process consisting in a simultaneous growth of the radiation field, the electron beam energy modulation, and the electron beam density modulation (also called microbunching). 
The FEL process starts with an initial radiation field, which in the standard FEL configuration (see next section for more details) is associated to the spontaneous undulator radiation generated by the electron beam. 
This initial field induces an energy modulation to the electron beam with a period equal to the radiation wavelength $\lambda$. The energy modulation is then converted to density modulation within the radiation wavelength $\lambda$. This microbunching results into an increase of the emitted radiation due to an increased degree of coherence. This, in turn, contributes again to enhance the energy modulation, and so on.   

The transverse oscillation of the electrons within the undulator allows the coupling between the electron beam and the co-propagating field. The energy transfer between an electron and the photon is proportional to $\vec{v}_\perp\vec{E}$, where $\vec{v}_\perp$ is the transverse velocity of the electron and $\vec{E}$ is the radiation field. 
A single electron will move either with or against the field line, losing or gaining energy with respect to the photon depending on the sign of the product $\vec{v}_\perp\vec{E}$.  
It can be demonstrated that, in the constructive interference condition, i.e. when the resonance condition (Eq.~\ref{eq:lam}) is fulfilled,  the direction of energy transfer remains constant over many undulator periods. 
For instance, after half undulator period the radiation field has slipped half wavelength, both velocity and field have changed sign and the direction of energy transfer stays the same.

The energy change of an electron with the longitudinal position in the undulator $z$ can be calculated as follows:
\begin{equation}
\frac{d\gamma}{dz} = \frac{-ef_cK}{2\gamma}\frac{E_0}{mc^2}\sin{\phi},
\label{eq:dg}
\end{equation}
where $f_c$ is the so-called coupling factor (with values around 0.9 for a planar undulator), $E_0$ is the initial peak radiation field, and $\phi$ is the phase within one wavelength. 
From the above expression it becomes clear that electrons with positive phase loose energy, while electrons with negative phases will gain energy. This will cause an energy modulation of the electron beam.

\begin{figure}[ht]
\begin{center}
\includegraphics[width=0.75\linewidth]{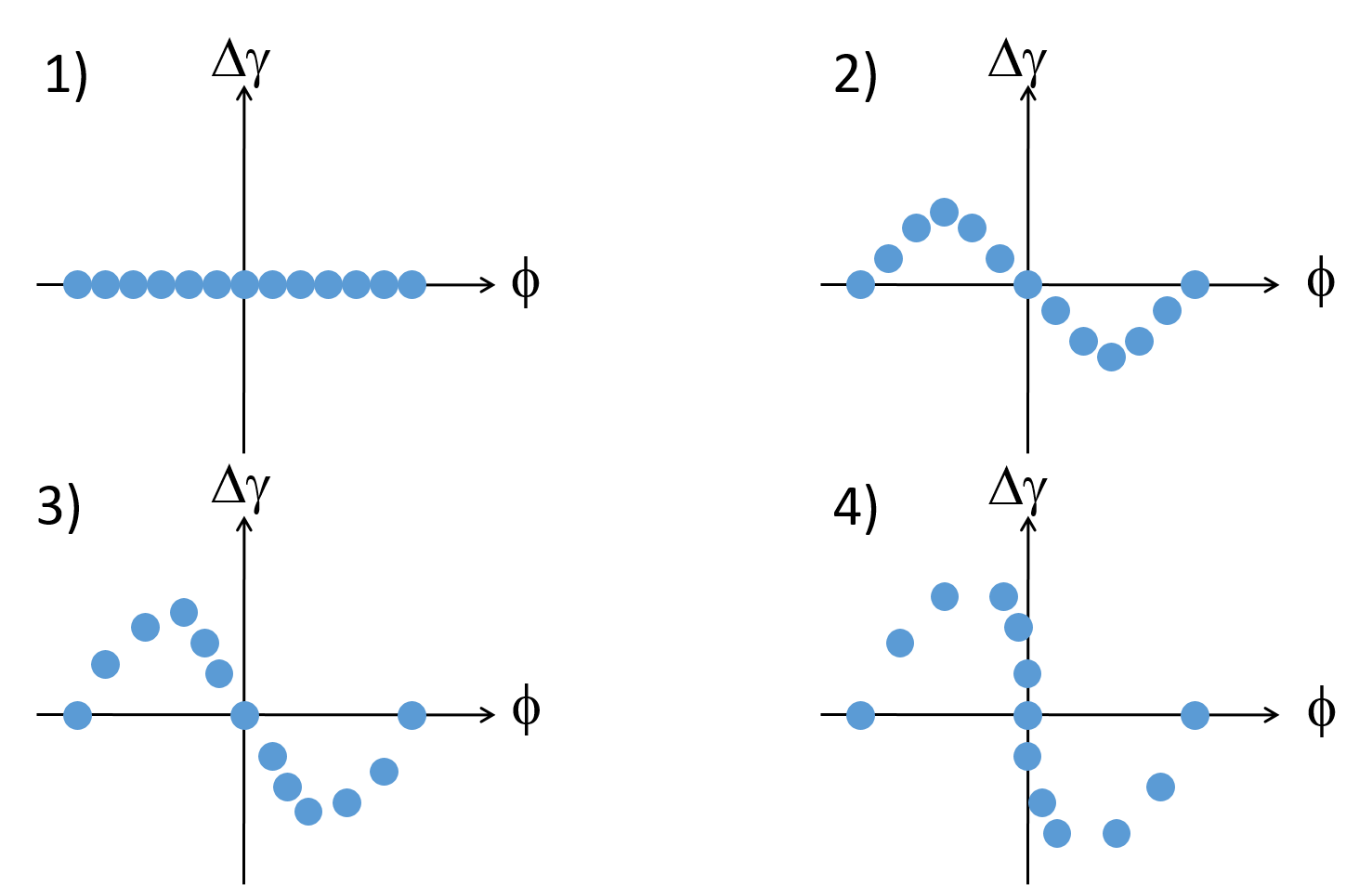}
\caption{Longitudinal phase space of the electrons (symbolized with blue circles) at different stages: 1) no energy or density modulation, 2) initial energy modulation, 3) enhanced energy modulation and initial density modulation, 4) enhanced density modulation (microbunching). See text for more details. Image courtesy: Sven Reiche.}
\label{fig:ubunching}
\end{center}
\end{figure}

Electrons gaining energy will move faster ($d\phi/dz > 0$), while electrons losing energy will fall back ($d\phi/dz < 0$). For small energy deviations $\Delta\gamma$ the following expression can be derived:
\begin{equation}
\frac{d\phi}{dz} = 2k_u\frac{\Delta\gamma}{\gamma_r},
\label{eq:d0}
\end{equation}
where $\gamma_r$ corresponds to the Lorentz factor that fulfills the resonance condition (Eq.~\ref{eq:lam}). 
Because of this effect the electrons will move together, they will be packed in a small fraction of the wavelength. In other words, the energy modulation will cause a density modulation or microbunching.  

With the help of Fig.~\ref{fig:ubunching} we can understand how the energy modulation and microbunching are generated. 
The radiation wavelength is normally much shorter than the bunch length, so at the beginning the electrons are randomly spread out over all phases (see Fig.~\ref{fig:ubunching}.1). Following Eq.~\ref{eq:dg} the electrons will start to modulate in energy (Fig.~\ref{fig:ubunching}.2). After being modulated in energy they start to also modulate in density according to Eq.~\ref{eq:d0} (Fig.~\ref{fig:ubunching}.3) until the beam is highly bunched (Fig.~\ref{fig:ubunching}.4).

\begin{figure}[ht]
\begin{center}
\includegraphics[width=0.75\linewidth]{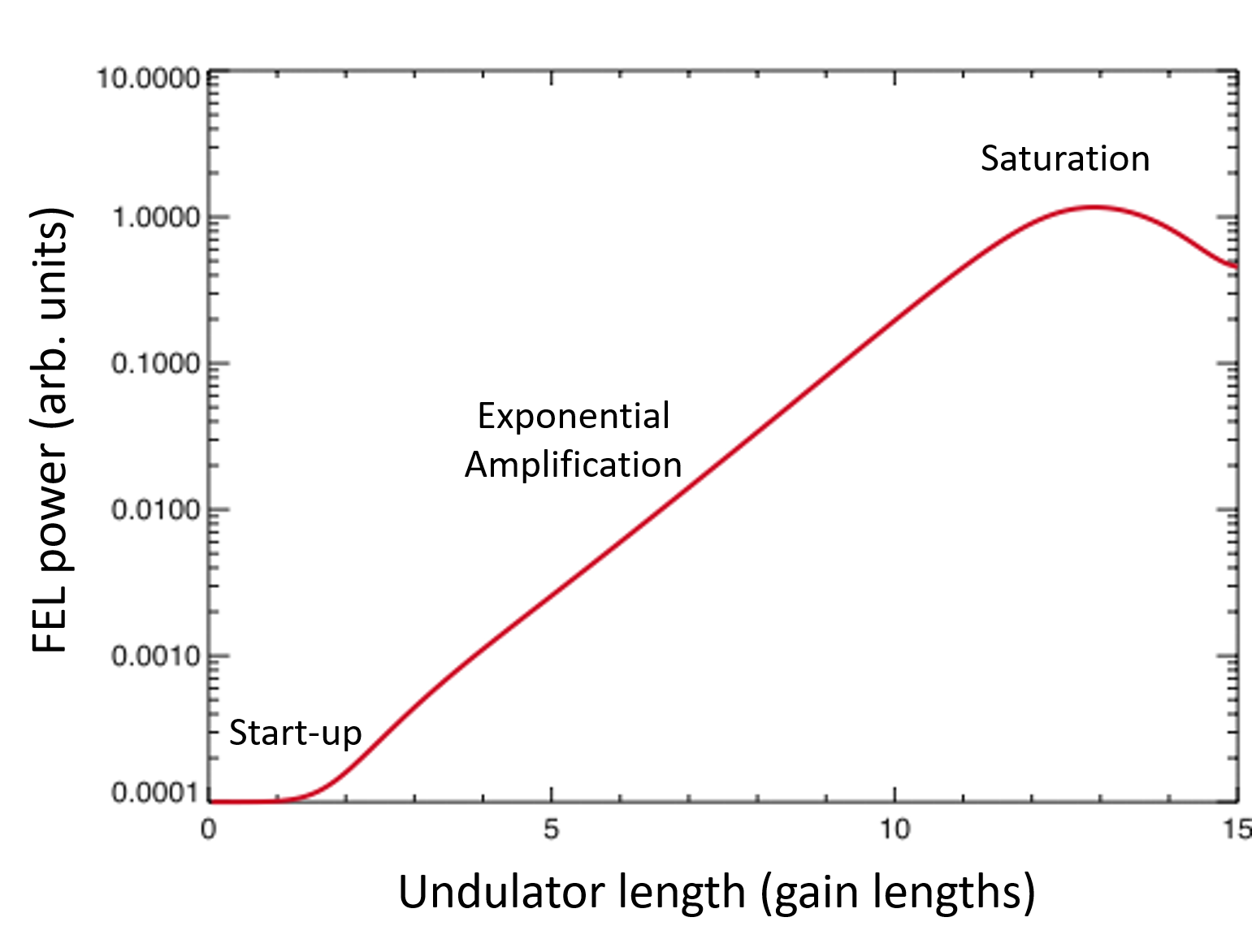}
\caption{FEL power along the undulator beamline. The vertical scale is logarithmic. See text for more details. Image courtesy: Sven Reiche.}
\label{fig:gain_curve}
\end{center}
\end{figure}

Figure~\ref{fig:gain_curve} shows the generic FEL amplification process; i.e. how the FEL radiation power evolves along the undulator line. The undulator length is displayed in units of gain length $L_G$, defined as the required distance so that the power gets multiplied by a factor of $e$ (in the exponential regime).
We distinguish different regimes. 
At the startup there is a lethargy where the power does not grow. This corresponds to the time until some microbunching has been formed. After the lethargy there is the  amplification regime in which the FEL power grows exponentially. The FEL process finishes with saturation. At this stage the microcbunching has reached its maximum. Beyond saturation there is a continuous exchange of energy between electron beam and radiation beam and the FEL power stays rather constant.

\subsubsection{FEL modes} 
\begin{figure}[ht]
\begin{center}
\includegraphics[width=0.8\linewidth]{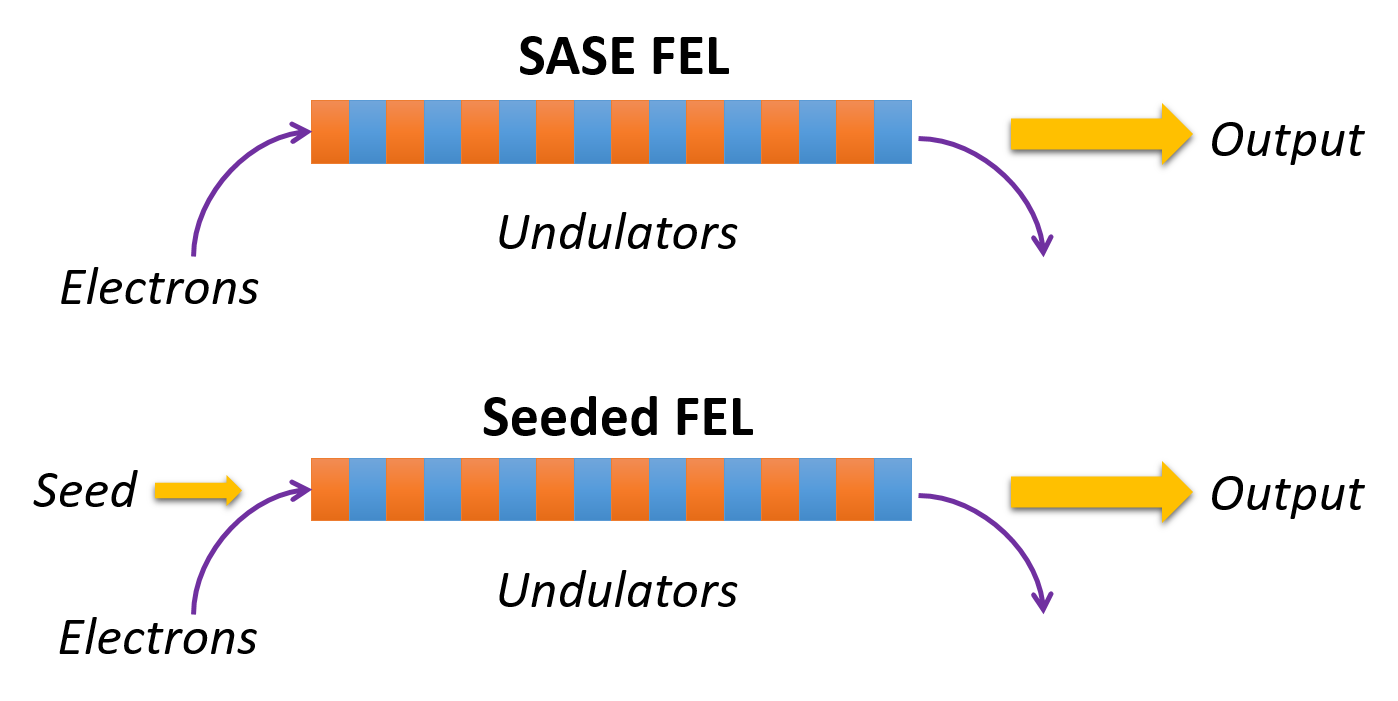}
\caption{Schematic layouts of SASE and seeded-FEL modes. Image courtesy: Sven Reiche.}
\label{fig:SASE_seed}
\end{center}
\end{figure}

The FEL can start either with the spontaneous undulator radiation generated by the electron beam or with an external radiation field. The first case, which is the standard FEL configuration, is called Self-Amplified Spontaneous Emission (SASE) FEL~\cite{Kon80,Bon84}. The second case is referred as seeded FEL. 
Figure~\ref{fig:SASE_seed} shows an schematic layout of the SASE and seeded FEL options. 

\subsubsubsection{SASE FELs}. 
The electron beam travels through the undulator to generate the output radiation. 
The FEL process starts with the broadband signal of the spontaneous radiation, also called shot noise.
SASE-FEL radiation is almost coherent in the transverse direction but not in the longitudinal one. 
Due to the lack of full longitudinal coherence, the spectrum and time profile contain several modes or spikes (a perfectly coherent pulse would only contain one mode in time and spectral domain).  
The FEL pulse duration $t_b$ corresponds to the spike width in the spectrum $1/t_b$, while the time duration of a single spike in time profile $t_c$ corresponds to the width of the full spectrum $1/t_c$. 
The bandwidth of the SASE-FEL radiation is of the order of the Pierce parameter (see later), with typical values for a X-ray FEL facility varying between $10^{-4}$ and $10^{-3}$. 
Most of the X-ray FEL facilities are based on the SASE mechanism due to its simplicity and the technical complexity associated to seeded FELs.

\subsubsubsection{Seeded FELs}
In this case, the input of the undulator is not only the electron beam but also a seed signal, which is amplified within the undulator. The power of the seed has to be larger than the shot noise power of the electron beam, otherwise the shot noise will be amplified. 
The output radiation will resemble the characteristics of the seed. For example, if the seed has a single mode in spectrum and time, the output radiation will also consist of a single mode. 
Seeding is used to improve the longitudinal coherence or to reduce the bandwidth of SASE-FELs. Fully coherent pulses can be obtained with the seeded-FEL process.  
There are various seeding methods. One possibility is to use the self-seeding mechanism~\cite{Fel97,Sal01}: first, standard SASE-FEL radiation is produced in a first undulator section, then a radiation monochromator reduces the bandwidth of the SASE pulses, finally the monochromatic signal is used as a seed in a second amplification stage. 
Another option is to employ external lasers. Here it is possible to seed directly with a high-harmonic generation source~\cite{Fer88} or to use more complicated layouts with modulators and chicanes such as in the high-gain harmonic generation~\cite{Yu91} or the echo-enabled harmonic generation~\cite{Stu09} schemes. Self-seeding has been proven to work for both soft and hard X-rays~\cite{Rat15,Ama12}, while external seeding using the HGHG and EEHG schemes have reached wavelengths of few nanometers, entering the soft X-ray regime~\cite{FERMI,Rib19}.

 \begin{figure}[ht]
\begin{center}
\includegraphics[width=0.95\linewidth]{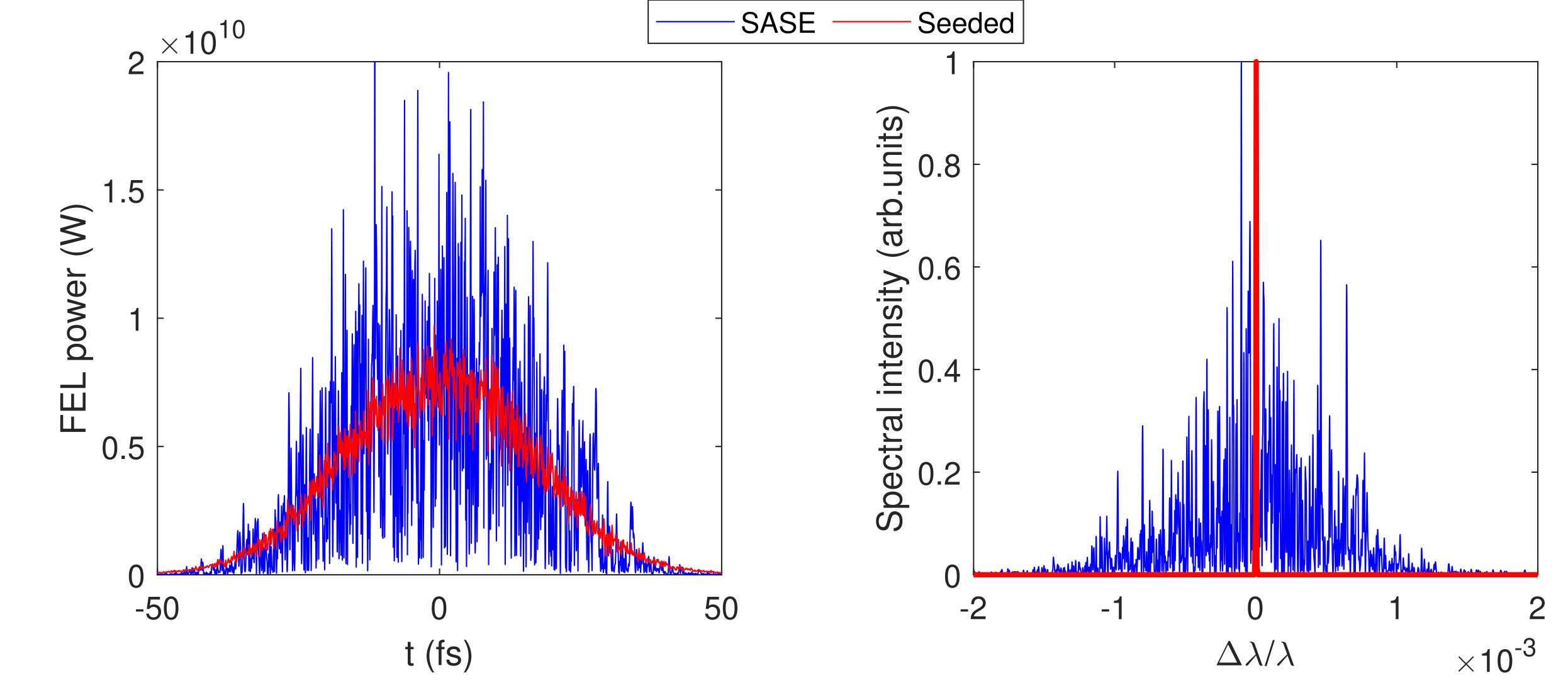}
\caption{Simulated power profile (left) and spectrum normalized to maximum intensity (right) of a SASE and a seeded FEL for a radiation wavelength of 0.1~nm. }
\label{fig:SASE}
\end{center}
\end{figure}

Figure~\ref{fig:SASE} shows a simulated power time profile and spectrum of a SASE and a seeded FEL at saturation for a radiation wavelength of 0.1~nm. In the seeded case, we consider a fully coherent initial field with a power of 1~MW. The rest of the simulation parameters are identical for both cases and correspond to the SwissFEL case~\cite{Pra20b}. The simulations have been done with the code Genesis~\cite{Gen}. As observed in the figure, the SASE FEL has a spiky structure with multiple modes in both time and spectral domains, while the seeded FEL contains a single spike in both domains and the bandwidth is much smaller than for the SASE case. 

\subsubsection{The Pierce parameter and the FEL properties}
The Pierce parameter $\rho$ is a fundamental quantity to characterize the FEL process and performance: 
\begin{equation}
\rho = \frac{1}{\gamma_r}\left[\left(\frac{f_c K}{4k_u\sigma_{t}}\right)^2\frac{I}{2I_A}\right]^{1/3},
\label{eq:rho}
\end{equation}
where $I$ is the electron beam peak current, $\sigma_{t}$ is the transverse electron beam size ($\sigma_{t}=\sqrt{\sigma_x\sigma_y}$), and $I_A$ is the Alfv\'en current ($\approx17$~kA). 
The Pierce parameter has typical values between $10^{-4}$ and $10^{-3}$ for X-ray FELs. 

From 1D theory, the FEL power and the FEL gain length can be obtained as~\cite{Bon84}: 
\begin{equation}
P=\rho P_e,
\label{eq:PFEL}
\end{equation}
\begin{equation}
L_g = \frac{\lambda_u}{4\pi\sqrt(3)\rho},
\label{eq:Lg}
\end{equation}
where $P_e$ is the power of the electron beam. 
The SASE cooperation length, defined as the radiation slippage over one gain length, and bandwidth (in terms of angular frequency $\omega$) can be obtained from the following expressions:
\begin{equation}
L_c = \frac{\lambda}{4\pi\rho},
\end{equation}
\begin{equation}
\frac{\Delta\omega}{\omega} = 2\rho.
\label{eq:dlam}
\end{equation}

\subsubsection{Electron beam requirements}
A better FEL performance, i.e. higher powers and shorter gain lengths, can be obtained for larger values of the Pierce parameter (see Eqs.~\ref{eq:PFEL} and \ref{eq:Lg}). This means, by looking at Eq.~\ref{eq:rho}, that, besides aiming for larger undulator fields $K$ (which require higher electron beam energies for the same wavelength, see Eq.~\ref{eq:lam}), the electron beam needs to have large peak currents and small transverse beam sizes. 

Moreover, we have to consider that only electrons within the FEL bandwidth (defined by Eq.~\ref{eq:dlam}) can contribute to the FEL gain. Consequently, the relative energy spread of the electron beam needs to be smaller than the $\rho$ parameter for an efficient FEL amplification; i.e.:
\begin{equation}
\frac{\sigma_\gamma}{\gamma}<\rho.
\end{equation}
For X-rays, the relative energy spread needs to be smaller than $10^{-3}$. 

Finally, we have to consider effects related to the transverse emittance. 
The transverse emittance of the electron beam is the area in the transverse phase-space $u-u'$, where $u$ refers to either $x$ or $y$, occupied by the fraction of the beam distribution determined by its second-order moments. It can be obtained as $\varepsilon_u = \sqrt{\sigma_u^2\sigma_u'^2-\sigma_{uu'}^2}$, where $\sigma_u'$ is the transverse divergence of the electron beam and $\sigma_{uu'}$ its transverse coupling term. 
On the photon side, the radiation emittance of the fundamental mode of the field is given by $\lambda/(4\pi)$~\cite{Sie86}. 
Electrons enclosed in this effective phase-space ellipse of the photons will radiate coherently into the fundamental radiation mode. 
Electrons outside this ellipse will emit into higher modes and will not contribute to the amplification of the fundamental mode.
This sets the following limit to the transverse emittance of the electron beam: 
\begin{equation}
\varepsilon_{u} \lesssim \frac{\lambda}{4\pi}
\end{equation}
In FELs it is customary to normalize the electron beam emittance with respect to the electron beam momentum $p$, resulting in the normalized emittance
\begin{equation}
\varepsilon_{n,u} = \frac{p}{mc}\varepsilon_{u} \approx \gamma\varepsilon_{u}.
\label{eq_en}
\end{equation}
The above limit can then be rewritten as
\begin{equation}
\frac{\varepsilon_{n,u}}{\gamma} \lesssim \frac{\lambda}{4\pi}.
\end{equation}
This condition implies that, for the X-ray regime, normalized emittances at the micrometer level or below are required.

In conclusion, the FEL process requires electron beams with large peak currents, small beam sizes, emittances and energy spreads. In other words, electron beams with high charge density in the 6D phase space are required for FELs.
For X-rays, the electron beam needs to have an energy at the GeV level, a peak current at the kA level, a relative energy spread between 0.01 and 0.1\%, normalized emittances at the micrometer level or below, and transverse beam sizes of a few tens of micrometers or smaller.
There are currently no electron sources that can produce such bunches directly. Instead they have to be accelerated and compressed. 
State-of-the-art X-ray FEL facilities employ linear accelerators to provide the drive electron beam, as we will see in the next chapter (circular accelerators are not capable of delivering the required electron beam properties, in particular not the high peak current).

\section{X-ray sources}

The figure of merit of many X-ray experiments and therefore used to compare different radiation sources is the brilliance or spectral brightness. 
We distinguish between peak brilliance, corresponding to the brilliance over the duration of the photon pulse and more appropriate for time-resolved experiments, and the average brilliance over time.  
The peak brilliance is defined as the number of photons emitted by the source per unit of time into a unit of solid angle, per unit of surface, and into a unit bandwidth of frequencies around the central one. 
The unit for peak brilliance is [photons / s / mm$^2$ / mrad$^2$ / 0.1\% (BW)]. 
The average brilliance is the peak brilliance multiplied by the photon pulse duration and the repetition rate of the source.

For both definitions, the brilliance $B$ is proportional to the number of photons $N_p$ and inversely proportional to a certain time interval $\Delta t$, the photon beam size $\sigma_r$, the photon beam divergence $\sigma_r'$, and the relative bandwidth $\Delta\lambda/\lambda$; i.e. 
\begin{equation}
B \propto \frac{N_p}{\Delta t\sigma_r\sigma_r'\frac{\Delta\lambda}{\lambda}}.
\end{equation}
The product of the photon beam size and divergence is at best $\lambda/4\pi$ for a fully transverse coherent mode, which is called diffraction emittance and corresponds to the emission from a single electron. 
The net product of the radiation beam size and divergence will be the convolution of this fully coherent mode with the electron beam distribution in size and angle, with the electron beam transverse emittance adding up to the diffraction emittance in squares. 
Therefore, the electron beam transverse emittance is a fundamental parameter significantly contributing to the brilliance via the actual photon beam size and divergence.
When the electron beam emittance is much smaller than the photon emittance, the mode is diffraction limited.

\begin{figure}[ht]
\begin{center}
\includegraphics[width=0.6\linewidth]{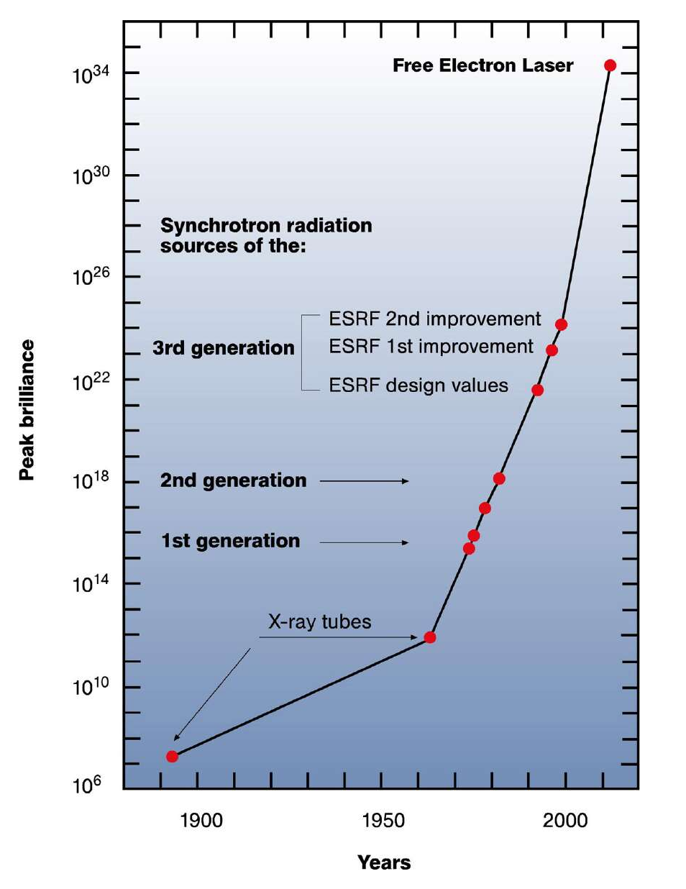}
\caption{Evolution of peak brilliance of X-ray sources. Image taken from~\cite{Mar06}}.
\label{fig:bri}
\end{center}
\end{figure}

Figure~\ref{fig:bri} shows the historical evolution of peak brilliance for X-rays. Synchrotron radiation machines of different generations are based on circular accelerators and can provide a peak brilliance up to $10^{25}$. X-ray FEL facilities are based on linacs, since only linacs can provide the required electron beam to drive the FEL process. Thanks to the increased pulse energies, shorter pulse durations, and lower bandwidths, FELs provide peak brilliances much higher than synchrotron facilities (e.g. about 10 orders of magnitude higher than third-generation synchrotrons). 
Nevertheless, since synchrotrons have higher repetition rates and longer pulses than X-ray FEL facilities, the difference in average brightness of synchrotrons and X-ray FELs is reduced to one or two orders of magnitude. 
We note that the values of the figure consider SASE-FELs for which the bandwidth is relatively large ($10^{-4}$ to $10^{-3}$). 
Seeded-FELs have demonstrated the possibility to reduce the bandwidth and therefore to increase the brilliance by an additional factor of 10-100~\cite{Ama12,Rat15}.

In the following we will give a brief overview of the history of synchrotron
light sources and FELs, list current facilities, and qualitatively describe
the layout as well as electron beam properties and dynamics of these machines.
We will do this from a general perspective, while using specific facilities
as examples, in particular the Swiss Light Source (SLS) and SwissFEL, the two
representatives of synchrotron and FEL facilities at the Paul Scherrer 
Institute in Switzerland.

\subsection{Synchrotron light machines}

\subsubsection{History and current facilities}
In the 1960s some electron storage rings designed and built for nuclear and subnuclear physics started to be used parasitically as sources of photons for experiments in atomic, molecular and solid state physics. These machines are nowadays referred to as first-generation light sources. 

Later, second-generation facilities were designed and optimized to serve exclusively as light sources using the synchrotron radiation emitted in bending magnets. 
Examples of these second-generation machines are BESSY I at BESSY (Germany), the two National Synchrotron Light Sources (NSLS and NSLS-II) at BNL (USA), and the Photon Factory at KEK (Japan).

In the 1990s third-generation light sources began to operate. These facilities were characterized by the extensive use of insertion devices (wigglers and undulators) and by the reduction of the beam emittance. These improvements
provided more brilliant, lower-bandwidth and tunable synchrotron radiation. Examples of this generation of sources are the European Synchrotron Radiation Facility or ESRF (France), Spring-8 (Japan), and the SLS at PSI (Switzerland).

A further improvement in the emittance by using multi-bend achromat (MBA) lattices (enabled by vacuum technology allowing smaller magnets) define the fourth-generation of synchrotron light sources. The brightness of such machines is one or two orders of magnitude larger than the third-generation light sources (not displayed in Fig.~\ref{fig:bri}). 
The fourth-generation machines have been pioneered by MAX-IV (Sweden), which started user operation in 2017~\cite{Tav18}. 
MAX-IV is followed by several other new projects and facility upgrades such as Sirius (Brazil)~\cite{Liu19} and the ESRF and SLS upgrades~\cite{Rai16,Str18}. 

There are presently around 50 synchrotron light sources around the world (operational or under construction) and about 15 new projects based on fourth-generation designs~\cite{lightsources}.  
Table~\ref{compSM} lists the start of operation and the main parameters (electron beam energy, circumference, electron beam current and horizontal emittance) of several synchrotron machines around the world~\cite{lightsources, ESRF-list}.

\begin{table}
\begin{center}
\caption{Start of operation and main parameters of several synchrotron radiation facilities around the globe~\cite{lightsources, ESRF-list}}.\label{compSM}
\begin{tabular}{ l c c c c c }
\hline\hline
\textbf{Name} & \textbf{Starting  } & \textbf{Energy } & \textbf{Circumference } & \textbf{Current } & \textbf{Emittance }  \\
& \textbf{operation year} & \textbf{ (GeV)} & \textbf{ (m)} & \textbf{ (mA)} & \textbf{ (nm)}  \\
\hline
Spring-8 (Japan) &1997 & 8 & 1436 & 100 & 3.0 \\
APS (USA) &1995 & 7 &  1104 &  100 & 3.0 \\
ESRF (France) &1994 & 6 &  844 & 200 &  3.8\\
Petra-III (Germany) &2009 & 6 &  2304 & 100 &  1\\
Diamond (UK) &2007 & 3 & 565 &  500 &  3.22\\
SSRF (China)  &2009 & 3 & 432 & 500 & 2.61 \\
Alba (Catalonia)  &2012 & 3 & 269  & 200  & 4.33 \\
NSLS-II (USA) &2015 & 3 &  792 &  500 &  0.55\\
TPS (Taiwan) &2016 & 3 & 518 & 500 &  1.6 \\
MAX-IV (Sweden) &2017 & 3 &  528 &  500 &  0.33\\
Sirius (Brazil) &2020 & 3 &  518 &  500 &  0.25\\
SLS (Switzerland) & 2001 & 2.4 & 288 & 400 & 5 \\
\hline\hline
\end{tabular}
\end{center}
\end{table}

\subsubsection{Layout}

\begin{figure}[ht]
\begin{center}
\includegraphics[width=0.75\linewidth]{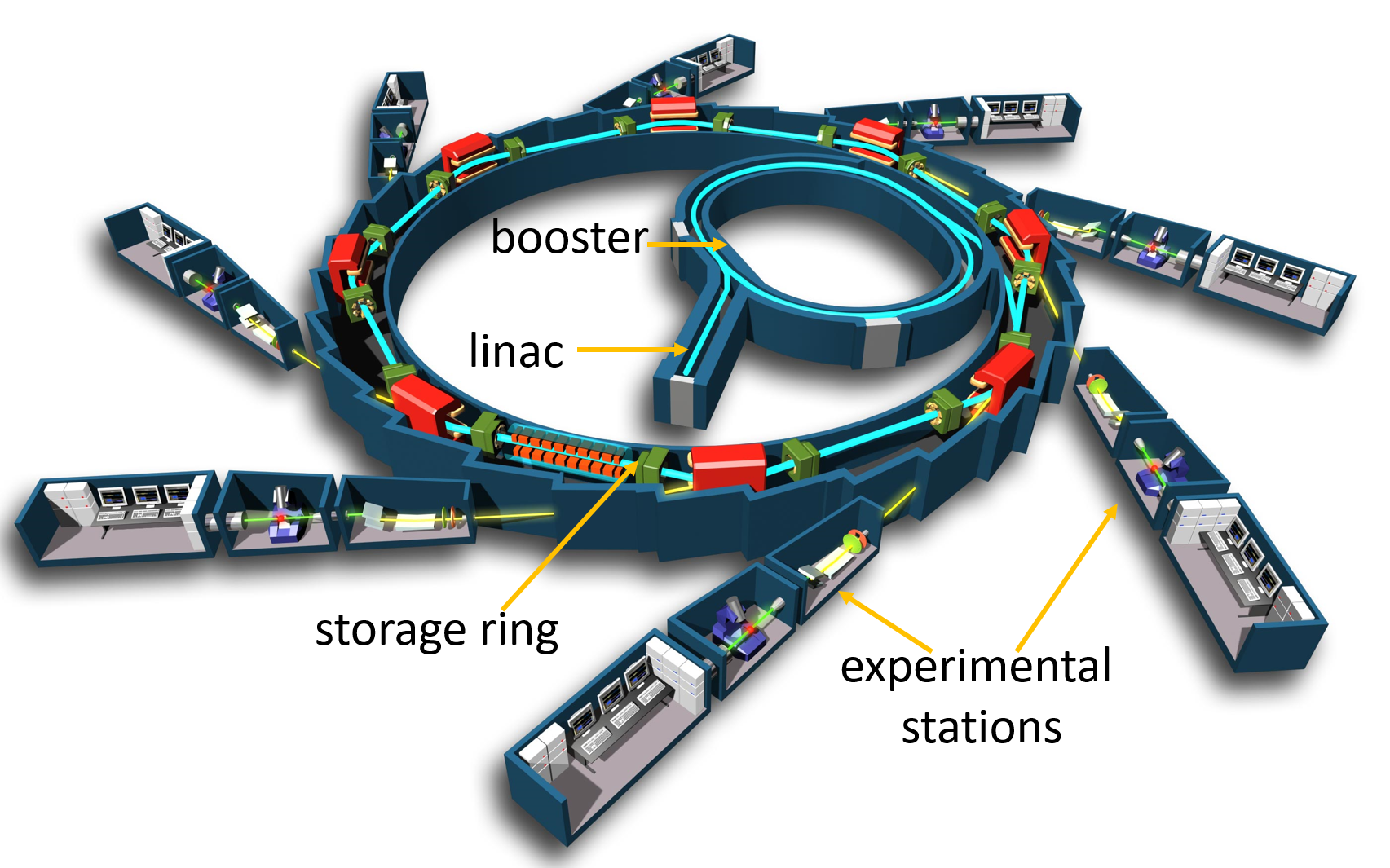}
\caption{Sketch of a typical layout of a synchrotron light facility. Original image from J.F.~Santarelli~\cite{San}.}
\label{fig:SM}
\end{center}
\end{figure}

Figure~\ref{fig:SM} shows a sketch of a typical layout of a synchrotron light source that includes four main parts: the linac, the booster, the storage ring, and the experimental stations. 
The linac consists of an electron source (a cathode), which generates the electrons, and RF sections, which accelerate the electron beam to an energy on the order of 100~MeV.
After the linac, the electrons are further accelerated with RF stations in the booster ring to its final energy of several GeV. 
The electrons are then injected and accumulated in the storage ring, where they produce synchrotron radiation, which is used in the experimental stations. 

The booster ring can be of the same size or smaller than the storage ring. The former case is the typical choice of modern machines due to several advantages: 
First, it implies a substantial saving on building space and shielding since the booster and the storage ring can be placed in the same tunnel. 
Second, the transverse emittance is improved since the total bending field can be distributed to a large number of small magnets (see next section for further details on the emittance). 
Third, it enables a cleaner injection into the storage ring and a simpler booster-ring transfer line. 
Fourth, there is a lower power consumption for the so-called top-up injection, in which the booster injects electrons periodically into the storage ring to keep the intensity of the circulating electron beam constant. Top up operation guarantees a constant heat load on all accelerator components, thereby improving machine stability. 

RF cavities in the storage ring resupply the energy lost by the electrons due to the emission of synchrotron radiation. 
The energy of the electrons in the storage ring amounts to several GeV. 
Higher electron beam energies allow higher photon energies and higher radiation fluxes but also require higher magnetic strengths (or larger circumference) and higher power consumption of the RF system to replenish the lost energy. 

The radiation is produced in bending magnets and insertion devices placed in the straight sections of the storage ring. 
The storage rings typically incorporate a variety of undulators covering a wide spectrum of synchrotron light between ultra-violet to hard X-rays. 
Synchrotron light machines operate with a relatively large number of experimental stations. 
As an example, in the SLS there are presently 16 beamlines in user operation.

\subsubsection{Electron beam properties and dynamics}
The emission of synchrotron radiation by the electrons in the storage ring gives rise to both radiation damping and quantum excitation. 
On the one hand, classical radiation damping is related to the continuous loss of energy of the electrons in bending magnets and wigglers or undulators and its compensation in the RF cavities. 
On the other hand, energy is also lost in discrete units or ``quanta,'' i.e., photons whose energy and time of emission vary randomly. This randomness introduces a type of noise or diffusion, causing a growth of the oscillation amplitudes of the electrons.
The distribution of the electrons in the storage ring, (i.e., emittance, pulse duration and energy spread), are determined by the balance or equilibrium between radiation damping and quantum excitation. 
More information on this topic can be found in Ref.~\cite{Wal94}. 

\begin{figure}[ht]
\begin{center}
\includegraphics[width=0.9\linewidth]{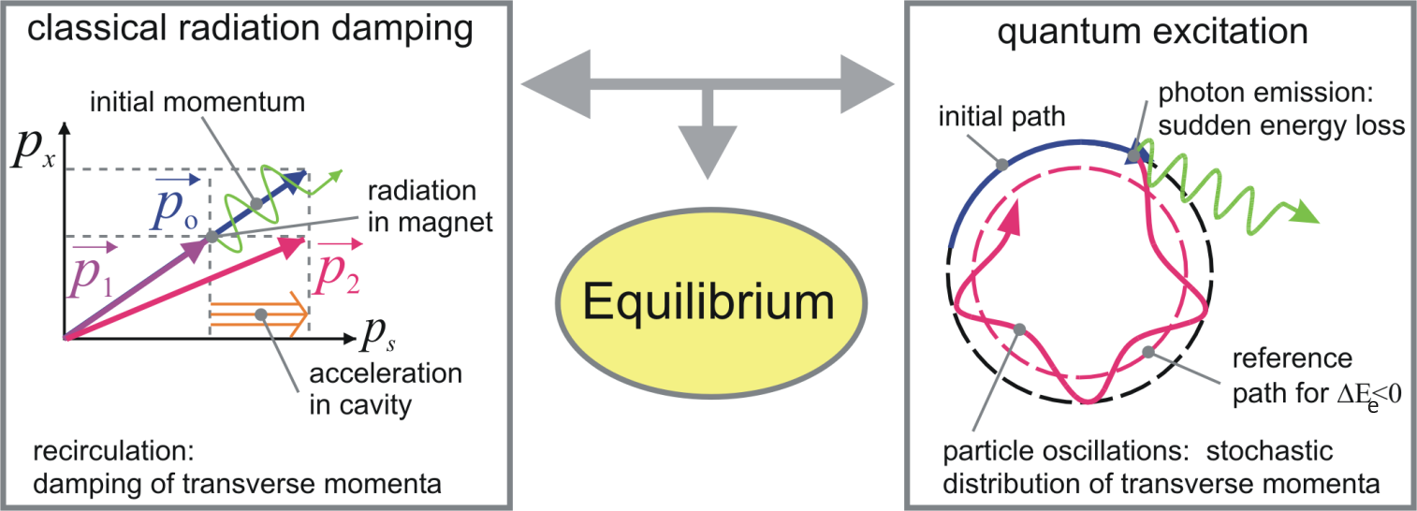}
\caption{Sketch to show the equilibrium that defines the electron beam horizontal emittance. Left: radiation damping. The continuous emission of synchrotron radiation causes a momentum reduction in both longitudinal and transverse directions. The RF cavities restore only the longitudinal momentum. As a result, the transverse momenta and therefore the emittance are reduced. Right: quantum excitation. Due to quantum emission of radiation, some electrons lose energy and follow different paths due to the finite dispersion in the lattice, causing the emittance to increase. Image courtesy of Andreas Streun.}
\label{fig:emit} 
\end{center}
\end{figure}

This equilibrium process is explained for horizontal momenta in the diagram shown in Fig.~\ref{fig:emit}. 
The emittance optimization will aim at maximizing radiation damping while minimizing quantum excitation. 
The maximization of the radiation damping is achieved by increasing the radiation power with high-field bending magnets and damping wigglers (where the radiation power increase has to be compensated with RF power). 
The effect of quantum excitation is enhanced by the transverse dispersion in those locations along the lattice where radiation is emitted, since the electrons will end up on a different trajectory after losing energy (the dispersion quantifies the momentum dependence of the electrons' trajectory). Consequently, the minimization of quantum excitation requires the minimization of the lattice dispersion wherever radiation is emitted. 
This can be done by using many short bending magnets (with small deflection angle) instead of few long devices (with large deflection angle).
While third-generation light sources typically feature double- or triple-bend achromat lattices (DBA and TBA, respectively), fourth-generation machines adopt multiple-bend achromats (MBA). 
The idea of using MBAs dates back to the 90s~\cite{Ein14}, but its implementation became only possible in recent years with the option to miniaturize the lattice components thanks to new developments in vacuum technology.  
A further possibility to reduce the effects from quantum excitation is to employ longitudinal gradient bending magnets, i.e., bending magnets for which the highest deflection field and  therefore emission is concentrated in a small region in the center.
The efficient use of longitudinal gradient bends in a periodic lattice cell requires in addition reverse bends to suppress the dispersion at these locations.
Third-generation synchrotron light sources have horizontal emittances on the order of a few nanometers, while fourth-generation machines can reduce the emittance below one nanometer (thanks to MBA lattices, damping wigglers and longitudinal gradient bends).

Until now we have talked about the horizontal emittance. 
In the vertical plane, the theoretical dispersion, in absence of errors, is zero, since the design dispersion arises only in the horizontal plane. 
In practice, however, the vertical emittance is dominated by magnet alignment errors. Third-generation machines typically achieve vertical emittances that are on the order of 1\% or less of the horizontal emittance.
As a notable example, a vertical emittance of 1~pm has been achieved at the SLS~\cite{Aib12}.

The bunch length and the energy spread of the electrons are also the result of the equilibrium between radiation damping and quantum excitation. Bunches are typically a few mm long, while relative energy spreads are usually at the 0.1\% level. The accumulated currents typically amount to a few hundred mA (averaged over a full turn). 
Synchrotrons operate at relatively high bunch frequencies -- as an example, the SLS works with a 500~MHz repetition rate, given by the choice of RF system. 
The design and operation optimization of synchrotron light sources includes addressing many further issues such as chromaticity correction and the optimization of beam lifetime and dynamic aperture, to name only a few.

\subsection{X-ray FEL machines}
\subsubsection{History and current facilities}

Both synchrotron light sources based on MBA lattices as well as X-ray FEL facilities are often referred to as fourth-generation light sources, which may be confusing at times.
The naming is justified insofar as both types of machines were developed around the same time and after the third generation of synchrotron light sources.

The first FEL facility to reach wavelengths in the soft X-ray regime was FLASH at DESY (Germany) in 2007~\cite{FLASH}. Thanks to its superconducting RF technology, FLASH operates with repetition rates at the MHz level. 
In 2009, the Linac Coherent Light Source (LCLS) at SLAC (USA) was the first FEL facility to produce radiation in the hard X-ray regime, with wavelengths at the {\AA}ngstrom level~\cite{LCLS}.
Two years after, in 2011, SACLA at Spring-8 (Japan) came into operation, producing hard X-ray FEL light with a much more compact accelerator than LCLS~\cite{SACLA}.
In 2013, FERMI at Elettra Sincrotrone Trieste (Italy) was the first fully seeded FEL going into operation delivering soft X-rays~\cite{FERMI}. 
In the last few years, three additional hard X-ray sources have started operation: PAL-XFEL at Pohang Accelerator Laboratory (South Korea)~\cite{Kan17}, the European XFEL at DESY (Germany)~\cite{EXFEL}, employing, like FLASH, superconducting RF structures to run at MHz frequencies, and SwissFEL at PSI (Switzerland), the most compact and cost-effective hard X-ray facility to date, driven by a high-brightness and relatively low-energy electron beam~\cite{Pra20b}.
Two more hard X-ray facilities are planned for the coming years: LCLS-II at SLAC (USA) and SHINE at SINAP (China), both of them adopting superconducting RF technology to be able to operate at MHz repetition rates. 
So far, all X-ray FEL facilities except FERMI are based on the self-amplified spontaneous emission (SASE) process. However, most of these facilities have the option to produce seeded radiation via the so-called self-seeding mechanism~\cite{Ama12, Rat15}.

Table~\ref{comp} lists the start of operation and the main parameters (electron beam energy, facility length, electron beam current and normalized emittance) of the hard X-ray FEL facilities operating to date~\cite{COMP,EXFEL,LCLS,Kan17,SACLA,Pra20b}.  
In contrast to Table~\ref{compSM} we quote the peak current (i.e., the current within a bunch) rather than the average current, and we give the emittance normalized with respect to beam energy, as is customary for FELs.
Furthermore, the reported emittance values correspond to the core of the bunch (slice emittance) for all facilities except SACLA, for which the emittance value refers to the full bunch (projected emittance).

\begin{table}
\begin{center}
\caption{Start of operation and main parameters of the hard X-ray FEL facilities presently operating worldwide~\cite{COMP,EXFEL,LCLS,Kan17,SACLA,Pra20b}}.\label{comp}
\begin{tabular}{ l c c c c c }
\hline\hline
\textbf{Name} & \textbf{Starting  } & \textbf{Energy } & \textbf{Length } & \textbf{Peak } & \textbf{Normalized }  \\
& \textbf{ operation year} & \textbf{ (GeV)} & \textbf{ (m)} & \textbf{current (kA)} & \textbf{emittance (nm)}  \\
\hline
European XFEL (Germany) & 2017 & 17.5 & 3400 & 5 & <600  \\
LCLS (USA)  &2009 & 14.3 & 3000 & 2.5--3.5 & 400 \\
PAL XFEL (Korea) & 2017 & 10 & 1100 & 2.5 & 550  \\
SACLA (Japan)  &2010 & 8.5 & 750 & >3 &  1000 \\
SwissFEL (Switzerland)  &2018 & 5.8 & 740 & 2 & 200 \\
\hline\hline
\end{tabular}
\end{center}
\end{table}

\subsubsection{Layout}

\begin{figure}[ht]
\begin{center}
\includegraphics[width=0.95\linewidth]{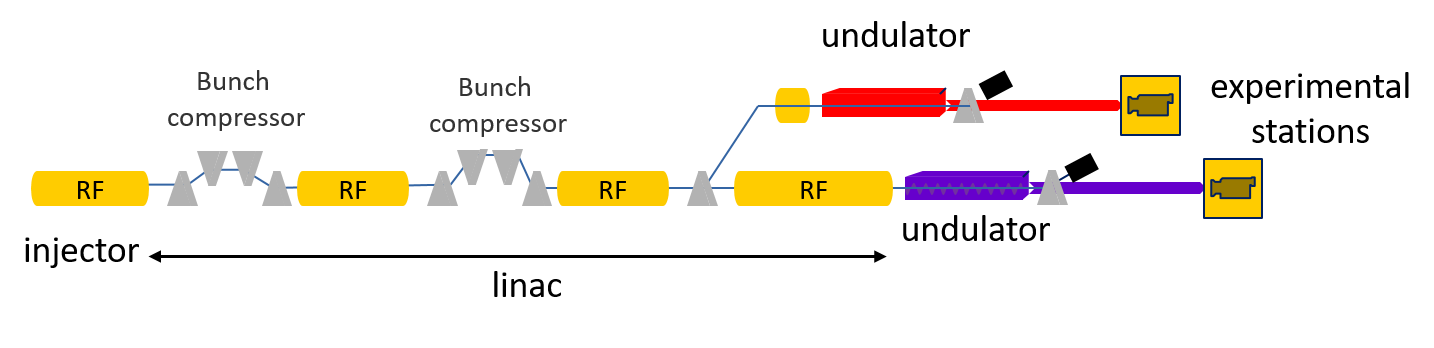}
\caption{Sketch of a typical X-ray FEL facility.}
\label{fig:SF}
\end{center}
\end{figure}

Figure~\ref{fig:SF} shows a schematic layout of a typical X-ray FEL. The facility can be separated into four main sections: the injector, where the high-brightness electrons are produced at relatively low energy and peak current; the linac, where the electrons are accelerated in RF sections and compressed in bunch compressors; the undulators, where the FEL process takes place; and the experimental stations where the produced FEL radiation is used. 

Most FELs use RF photoinjectors to generate high-brightness electron beams with energies of 5--10~MeV and peak currents of 10--20~A. 
Electrons are released via the photo-electric effect by a laser of a proper wavelength impinging on a photocathode mounted inside an RF gun. The cathode material is normally copper, although certain facilities such as European XFEL and SwissFEL operate with a cesium-telluride coating to profit from its higher quantum efficiency. 

In the linac the electrons are accelerated to GeV energies and compressed to kA peak currents. 
Bunch compression is normally achieved in two or more sections within the linac. 
The accelerating frequency is in the GHz range. For instance, LCLS employs S-band technology (about 3~GHz) while SACLA and SwissFEL use C-band frequency (about 6~GHz). Higher frequencies permit acceleration of the beam over shorter distances and consequently result in a more compact linac. 
The linac can consist of normal-conducting RF cavities (``warm'' technology) or superconducting ones (``cold'' technology). In the latter case, many more pulses per second can be accelerated (e.g., 27\,000 pulses per second at the European XFEL with superconducting RF compared to 100 pulses per second at SwissFEL with normal-conducting RF).

The undulator beamline consists of several undulator modules, each of them with a length of a few meters and a period of a few centimeters. The space between two undulator modules, typically measuring around 1~m, is used for focusing the beam with quadrupole magnets as well as for diagnostic purposes. 
At the hard X-ray beamline of SwissFEL, for example, the undulator consists of 13 units, each of them with an undulator period of 15~mm, an undulator field parameter $K$ tunable between 1 and 1.8, and a total length of 4~m. The interundulator section occupies 0.75~m. 
X-ray FEL facilities run one or few undulator beamlines, each of them serving multiple experimental stations.

\subsubsection{Electron beam properties and dynamics} 

In linac-based X-ray FELs, which are single-pass machines, the electron beam properties are not the result of an equilibrium due to the emission of synchrotron radiation as in circular accelerators. 
Rather, the properties are defined at the electron source, i.e., the RF photoinjector. The normalized emittances will ideally be preserved through the linac according to Liouville's theorem. However, Liouville's theorem will not be fulfilled and the beam emittances will be increased in the presence of some deteriorating stochastic effects such as intra-beam scattering, non-linear space-charge forces or emission of coherent synchrotron radiation. 
The main goal in designing and operating an FEL facility is to produce a high-brightness electron beam at the injector, accelerate it and compress it in the linac while preserving its quality as much as possible. 

The RF photoinjector produces electron beams with typical energies between 5 and 10 MeV, peak currents in the range 10--20~A, low normalized emittances (micrometer or below) and energy spreads at the level of a few keV.
In the case of SwissFEL, the achieved injector parameters are a beam energy of 7.1~MeV, a peak current of 20~A, a normalized emittance of around 200~nm, and an energy spread of around 10~keV. 

The emittance of the source is determined by three different components: intrinsic emittance of the cathode (mostly depending on the laser size at the cathode), space-charge forces, and RF field effects in the gun. 
The gun gradient is set as high as possible with available technology. 
In the SwissFEL example, the maximum RF field is 100~MV/m, resulting in the energy of 7.1~MeV at the gun exit quoted earlier. 
The gun is equipped with a solenoid magnet used to focus the electron beam. 
The laser spot size and the field generated by the gun solenoid are optimized to counteract the contributions of the intrinsic emittance (smaller for smaller spot size) and space charge (smaller for larger spot size) to the final emittance. 
This procedure is often referred to as emittance compensation~\cite{Ser97}. 

\begin{figure}[ht]
\begin{center}
\includegraphics[width=0.75\linewidth]{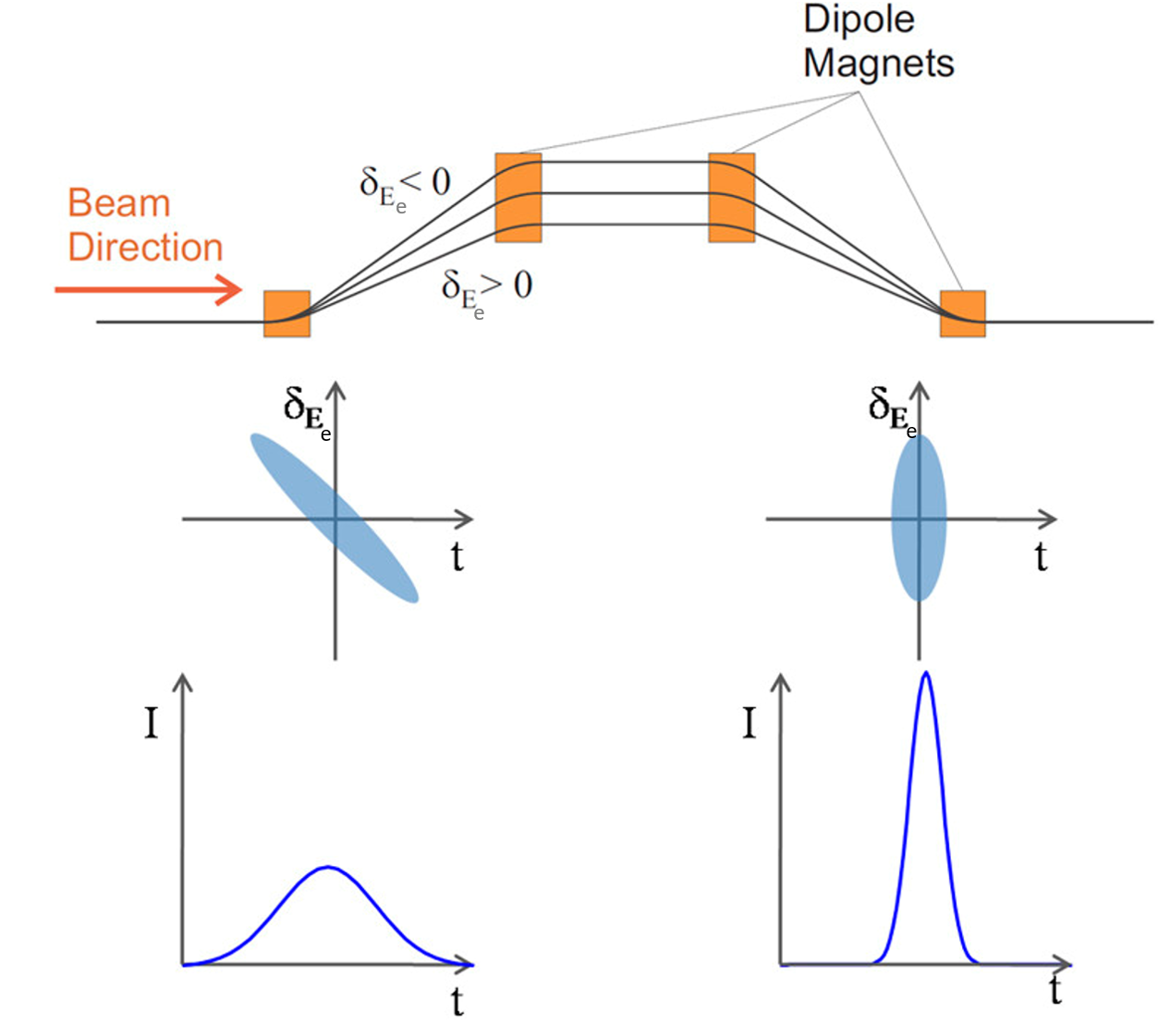}
\caption{Principle of bunch compression. Top: trajectories of electrons with different energies ($\delta_{E_e}$ indicates relative energy variation) through a magnetic chicane of four dipole magnets. Center: longitudinal phase-space of the electron beam at the entrance (left) and the exit of the magnetic chicane. Bottom: current profile at the entrance (left) and exit (right) of the chicane. Image taken from~\cite{BC_SF}.}
\label{fig:BC}
\end{center}
\end{figure}

Bunch compression follows a principle similar to FEL microbunching and is a two-stage process: First, the accelerating phases of some RF structures are tuned to imprint an energy chirp along the electron beam (i.e., a strong correlation between the energy and the longitudinal coordinate of the electrons along the bunch). Second, the beam is transported through a magnetic chicane, typically consisting of four bending magnets. Thanks to the energy dependence of the path along the chicane, the electron beam is compressed after the chicane. Figure~\ref{fig:BC} illustrates how bunch compression works. 

Bunch compression should be performed at suitable beam energies and currents to avoid beams with too high charge density, which would deteriorate the electron beam quality (in particular the emittance). 
Bunch compression is normally achieved in two or three stages along the linac. 
The first bunch compressor is typically at an energy of a few hundreds of MeV (300~MeV in the SwissFEL case) and the second or third when the energy already exceeds 1~GeV (2.1~GeV at SwissFEL). Each stage compresses the beam by a factor of around 10, resulting in a total bunch compression factor of around 100. 
At SwissFEL, the beam is compressed in the first bunch compressor to increase the peak current from 20~A to about 150~A, while the second bunch compressor further compresses the bunch to its final peak current of 2--3~kA. 
In general, high charge densities give rise to strong electromagnetic fields generated by the electron bunches, which in turn affect the electrons within a bunch. 
Examples of such interactions include coherent synchrotron radiation in bunch compression chicanes, wakefields, and space-charge forces.

\subsection{Comparison and conclusion} 

To conclude let's compare the electron beam properties of synchrotron and X-ray FEL facilities. 
First, the transverse electron beam emittances of X-ray FELs are somewhat similar with respect to the last generation of synchrotrons. For instance, the normalized emittance at SwissFEL is 200~nm (see Table~\ref{comp}), which corresponds to a (non-normalized) emittance of approximately 18~pm for a beam energy of 5.8~GeV (see Eq.~\ref{eq_en}). 
MAX-IV has a horizontal emittance of 330~pm (see Table~\ref{compSM}), 18 times larger than the one at SwissFEL. 
In the vertical plane, however, the emittance in synchrotrons is better. For example, the vertical emittance at SLS, as mentioned earlier, is as low as 1~pm, 18 times smaller than at SwissFEL. 
We can therefore conclude that the four-dimensional transverse emittances are essentially equivalent in both types of machines, with X-ray FELs featuring rounder beams and synchrotrons producing asymmetric beams with large horizontal but smaller vertical dimensions.

The energy spreads are significantly smaller in X-ray FELs than in synchrotrons ($10^{-4}$ versus $10^{-3}$). 
The main difference between the two types of machines is, however, the current and pulse duration: while synchrotrons have peak currents per bunch on the order of 10~A and pulse durations of tens of picoseconds, X-ray FELs provide kA peak currents and pulse durations at the femtosecond level.
Considering that the pulse duration of the produced radiation is normally equivalent to the one of the driving electron beam, X-ray FELs provide much shorter radiation pulses suitable for studying, e.g., chemical reactions with the appropriate time resolution.

The FEL process generates radiation with significantly higher power levels and lower bandwidths than the light emitted in synchrotron machines: while X-ray FELs typically provide pulse energies at the mJ level and relative bandwidths of $10^{-4}$, synchrotrons give nJ pulse energies and much larger relative bandwidths. 
All in all, the peak brilliance of the produced radiation of X-ray FELs is much higher than the one obtained at synchrotron facilities (as shown in Fig.~\ref{fig:bri}).

Nevertheless, synchrotrons do have their advantages: 
First, they provide much higher repetition rates than X-ray FELs, which is beneficial for many experiments. For instance, SLS works at 500~MHz while most of X-ray FELs operate at the 0.1~kHz level or at most up to about 1~MHz (if operating with superconducting RF technology).
Factoring in the higher repetition rates and the longer pulse durations at synchrotrons, the differences in average brilliance become much less pronounced than those in peak brilliance (shown in Fig.~\ref{fig:bri}). 
Another advantage of synchrotrons has to do with user accessibility. 
Since there are many more synchrotrons (more than 50 worldwide) than X-ray FELs (5 to 10), and since each synchrotron has more beamlines (up to about 50) than X-ray FELs (up to about 10), the access to synchrotrons is much easier than to X-ray FELs.

To summarize, both synchrotrons and X-ray FELs are invaluable research tools capable of observing matter with spatial resolution at the atomic level. 
X-ray FELs offer higher peak brilliance, higher pulse energies and a better time resolution suitable to study the dynamics of processes occurring at the atomic level, but synchrotrons have a higher repetition rate and provide easier access to the scientific community.

\section*{Acknowledgments}
We would like to thank Sven Reiche and Thomas Schietinger for valuable comments and discussions that help to improve the quality and the language of the document.

\end{document}